\documentstyle[12pt]{article}
\begin{document}

\title{ Ginzburg-Landau equation for steps on creep curve}
\author{ Mulugeta Bekele$^{1,}$\thanks {On leave from Department of Physics, 
Addis Ababa University, Addis Ababa, Ethiopia} and G.
Ananthakrishna$^2$\\ $^1$ Department of Physics, $^2$ Materials
Research Centre\\ Indian Institute of Science, Bangalore 560
012, India\\}
\date{}
\maketitle
\begin{abstract}
We consider a model proposed earlier by us for describing a form
of plastic instability found in creep experiments .  The model
consists of three types of dislocations and some transformations
between them. The model is known to reproduce a number of
experimentally observed features.  The mechanism for the
phenomenon has been shown to be Hopf bifurcation with respect to
physically relevant drive parameters.  Here, we present a
mathematical analysis of adiabatically eliminating the fast mode
and obtaining a Ginzburg-Landau equation for the slow modes
associated with the steps on creep curve. The transition to
the instability region is found to be one of subcritical bifurcation
over most of the interval of one of the parameters while 
supercritical bifurcation is found in a narrow mid-range of the 
parameter. This result is consistent with experiments. The
dependence of the amplitude and the period of strain jumps on stress and
temperature derived from the Ginzburg-Landau equation are also 
consistent with experiments. On the basis of detailed numerical 
solution via power series expansion, we show that high order 
nonlinearities control a large portion of the subcritical domain.
\end{abstract}

\newpage

\section{ Introduction} 
Instabilities in plastic flow has been an object of attention
for a long time in metallurgical literature. Conceptually
simplest form of this [Lubahn \& Felgar, 1961; Hall, 1970]
manifests when the material is subjected to a creep test wherein
a force is applied and the response in the form of elongation of
the material is measured, which under normal conditions, is a
smooth strain-time curve.  However, under certain metallurgical
conditions, one sees steps on creep curve suggesting a form of
instability [Lubahn \& Felgar, 1961; Hall, 1970; Da Silveira \&
Monteiro, 1979].  A better known form of the instability having
the same physical origin, but conceptually more difficult,
appears when the material is deformed under tensile deformation
[Bodner \& Rosen, 1967; Brindley \& Worthington, 1970; Penning,
1972].  Here, the material is subjected to a predetermined
response namely a constant rate of deformation, and the force or
the stress developed in the sample is sought to be measured.
Even in this case, one finds a smooth stress-strain curve under
normal conditions. However, when the system is in the regime of
instability, for some values of material parameters the
stress-strain curve exhibits multiple load drops.  Each of the
load drops is related to the formation and propagation of
dislocation bands [Chihab {\it et al}, 1987]. It is in the latter
type of testing where plastic instability manifests much more
easily than in the former. The phenomenon is referred to as the
Portevin-Le Chatelier (PLC) effect or the jerky flow and is seen
in several metals such as commercial aluminium, brass, alloys of
aluminium and magnesium [Brindley \& Worthington, 1970]. It is
observed only in a window of strain rates and temperature. In
contrast, the phenomenon of steps on creep curve, which is the
subject of the present discussion, is seen in few
instances [Da Silveira \& Monteiro, 1979; Zagarukuyko {\it et
al}, 1977; Stejskalova {\it et al }, 1981 ]. The reason
attributed to this is that it is difficult to obtain proper control
on metallurgical parameters wherein this form of instability can be
observed. There is one more form of
instability which manifests under constant stress test where
similar strain jumps are seen as a function of time.
The origin of instability in all these three modes of
testing is known to be the same.( For a current status of 
both experiments and theory see Kubin {\it et al}, 1993). Eventhough steps
on creep curve are seen in a limited number of experiments, it
is often straight forward to translate many of the experimental 
results of constant strain rate test into that of creep test. 
It is generally agreed that the microscopic origin
of the instabilities arises due to the interaction of
dislocations with mobile point defects and is referred to as
dynamic strain ageing. This leads to negative strain rate
characteristic. The basic idea was formulated by Cottrell [1953]
few decades ago.

The early phenomenological models are all static since they do
not deal with time development. In contrast, methodology of
dynamical systems addresses precisely this aspect.  Until about
a decade and a half ago, there were no models which looked at 
the problem from the point of view of bifurcation theory. 
An attempt to understand the problem in the above
perspective was first made by our group several years ago
[Ananthakrishna \& Sahoo, 1981b; Ananthakrishna \& Valsakumar,
1982, 1983; Valsakumar \& Ananthakrishna, 1983]. In a series of
papers starting from an extended Fokker-Planck equation for the
distribution function of the velocity of dislocation segments,
and then splitting this into a mobile and an immobile component,
 we arrived at a model which
consisted of three types of dislocations and some
transformations between them [Ananthakrishna \& Sahoo, 1981a,
1981b; Sahoo \& Ananthakrishna, 1982; 
Valsakumar \& Ananthakrishna, 1983]. The basic idea
could be summarized by stating that the phenomena is due to
a Hopf bifurcation resulting from nonlinear interactions between 
three different types of
dislocations, suggesting a new mathematical mechanism for the
instability.  Eventhough the spatial inhomogeneous structure
was ignored and only the temporal oscillatory state was sought
to be described, the model and its extensions to the case of
constant strain rate test, proved to be very successful in that
it could explain most of the experimentally observed features
(for critical reviews see Kubin \& Lepinoux, 1988 and
Kubin {\it et al}, 1993).
{\it For instance, the existence of negative strain rate sensitivity
in a regime of strain rates, comes out naturally as a consequence of 
the Hopf bifurcation} [Ananthakrishna \&
Valsakumar, 1982]. It
must be emphasized, that this feature has been measured in most
experiments on the PLC effect, and  is {\it assumed} in all other 
theoretical models 
[ Cottrell, 1953; Bodner \& Rosen, 1967; Penning, 1972; 
Kubin and Estrin 1990; see also Kubin \& Martin, 1988; 
Martin \& Kubin, 1992; Kubin {\it et al} 1993].
Other experimentally observed features  such as the 
existence of bounds on strain rate for the PLC
effect to occur,  the existence of critical strain and its dependence
on applied strain rate, the dependence of the amplitude on 
the strain etc., also follow. {\it One other
important prediction which is direct consequence of the dynamical basis 
of the model is the existence of chaotic
stress drops in a window of strain rates} [Ananthakrishna \&
Valsakumar, 1983; Ananthakrishna \& John, 1990]. 
{\it Recently this prediction has been verified which in turn
implies that only a few degrees of freedom are required 
for a dynamical description of
the phenomenon.  This offers
justification for the use of only a few degrees of freedom for
the description of the temporal aspect, eventhough the system is
spatially extended.} 
 [Ananthakrishna {\it et al}., 1995;
Ananthakrishna \& Noronha, 1995; Quaouire \& Fressengeas, 1995;
Venkadesan {\it et al} 1995; Noronha {\it et al} 1996 \& 1997]. 
(Note that a
spatially extended system implies infinite degrees of freedom.)
Description of the phenomenon which includes the initiation and
propagation of the bands during the PLC effect has also been
recently attempted [Ananthakrishna, 1993].

Since the introduction of bifurcation theory into this field
several years ago by our group, there has been a resurgence
of interest in plastic instabilities in the light of
introduction of new methodology borrowed from the theory of
dynamical systems. This has further helped to obtain new
insights hitherto not possible [Kubin \& Martin, 1988; Estrin \&
Kubin, 1989; Kubin \& Estrin, 1990; see also Kubin \& Martin, 1988;
Martin \& Kubin, 1992; Kubin, 1993; Ananthakrishna {\it et al}., 
1995a \&b and references therein].
One of the aims of such theories is to be able to relate the
microscopic dislocation mechanisms to the measurable macroscopic
quantities. However, in the process, we feel that finer aspects
of dynamical systems have been glossed over in this field. For
instance, one often finds that casual remarks are made about
fast and slow modes without actually going through the procedure
of demonstrating the existence of such modes and eliminating the
fast modes in favour of the slow ones [Aifantis, 1988; Hahner \&
Kubin, 1992; Hahner, 1993]. In addition, under the adiabatic
elimination, the slow modes are complicated functions of the original modes.
Yet, hand waiving arguments have been used in building models
which we believe are technically suspect.

The purpose of this article is at least three fold. First, we wish
to illustrate the power and utility of dynamical methods 
to the study of dynamical aspects of the model. We demonstrate the 
technical aspects of adiabatic elimination
(largely addressed to metallurgists working in the area of plastic
instabilities) and then derive the equation for the slow modes having
the form of time-dependent Ginzburg-Landau (TDGL) equation.
Second, we wish to compare the results of the dependence of the 
amplitude and the period of the strain jumps on stress and temperature
with experiments on creep curve as also with the results translated from the 
constant strain rate case
to the creep case. This will allow us to relate the theoretically 
introduced parameters to the macroscopically measured quantities.
Third, we wish to compare these results and the detailed numerical results
via power series expansion with the approximate solutions 
for the limit cycles obtained earlier [Ananthakrishna
\& Sahoo, 1981b; Valsakumar \& Ananthakrishna, 1983].

The principal results of this work are as follows.  We show that
the derived quintic TDGL equation for the slow modes is valid 
in the mid-range of the instability domain.
Within this range, there is a narrow sub-range of the parameter
where the system exhibits a supercritical bifurcation, beyond
which subcritical bifurcation is exhibited.  The results of this
calculations on the amplitude and period of the strain jumps are
compared with experiments. In addition, analytical work
based on the TDGL equation and a detailed comparative numerical
study via the power series solution shows that the model
exhibits unusual properties in a certain domain of the
parameter.  The study further shows that very high order
nonlinearities actually govern the nature of the subcritical
bifurcation beyond the domain handled by the quintic TDGL
equation.

The plan of the paper is as follows.  In what follows (section
2) we present a brief summary of the model.  In section 3, the
technique of extracting the TDGL equation using reductive
perturbative approach is applied to the model. In section 4,
approximate limit cycle solution obtained through the TDGL
equation is compared with experiments and 
with the numerical solution of the model as
well as  earlier studies on the model.  Section 5 contains
summary and discussion.

\section{ A Model for Steps on Creep Curve}

We start with a brief summary of the model.  The basic idea of
the model is that instabilities in plastic flow are a
consequence of nonlinear interactions between different types of
dislocation populations. Spatial dependence is ignored and
only temporal aspects are described, the idea being that the few
degrees of freedom  used for describing the phenomenon
(dislocation populations) correspond to the
collective modes of dislocations in the spatially extended
system. As mentioned in the introduction, this finds support from
the recent experimental verification of chaos in the PLC effect.
The details of the model
can be found in the original references [Ananthakrishna \&
Sahoo, 1981b; Valsakumar \& Ananthakrishna, 1983]. The model
consists of mobile dislocations $m$ and immobile
dislocations $im$ and another type which mimics the Cottrell's
type $i$, which are dislocations with clouds of solute atoms.
Let the corresponding densities be $N_m$, $N_{im}$ and $N_i$,
respectively.  The basic dislocation mechanisms included are the
following: (a) production of dislocations by cross glide 
(\( m \stackrel{\theta V_m}{\rightarrow} m + m \) , 
$\theta$ is the cross glide coefficient, $V_m$ is the velocity of
mobile dislocations; $V_m = V_0
(\sigma_a/\sigma_0)^m$ [Alexander, 1986] where $\sigma_a$ is 
the applied stress), (b) immobilization of two mobile dislocations 
(\( m + m \stackrel{\beta}{\rightarrow} im + im \), $\beta$ is the rate at
which two mobile dislocations get immobilized) and annihilation of two 
mobile dislocations 
(\( m + m \stackrel{\beta^{\prime}}{\rightarrow} 0 \)), and annihilation of a
mobile with an immobile dislocation 
(\( m + im \stackrel{\bar{\beta}}{\rightarrow} 0 \)),  (c) once a mobile
dislocation is immobilized, it can be reactivated by
athermal or thermal means 
(\( i \stackrel{\gamma}{\rightarrow} m \)), (d) lastly, 
a mobile dislocation can
acquire a cloud of solute atoms and move with them. We consider
such dislocations as distinct from both the mobile as well as
the immobile and denote them $i$. This process is represented by
\( m \stackrel{\alpha_m}{\rightarrow} i \), $\alpha_m$ being a
function of the concentration of solute atoms.  However, as the
solute atoms gather progressively around dislocations, these
dislocations eventually will be immobilized 
(\( i\stackrel{\alpha_i}{\rightarrow} im \)). ( Note that in principle,
$\beta$, $\beta^{\prime}$ and $\bar{\beta}$ are different.) 
These mechanisms lead to the rate equations for the densities of dislocations:
\begin{eqnarray}
\dot{N}_m & = & \theta V_m N_m - (\beta + \beta^{\prime}) N_m^2 - \bar{\beta} N_m N_{im}
+\gamma N_{im} -\alpha_m N_m\,,
\\
\dot{N}_{im} & = &\beta N^2_m - \bar{\beta} N_{im} N_m - \gamma N_{im} + \alpha_i N_i,
\\
\dot{N}_i & = & \alpha_m N_m - \alpha_i N_i.
\end{eqnarray}

\noindent
(For the sake of simplicity, we take $\beta^{\prime} = 0$ 
and $\bar{\beta} = \beta$.)  
One can give a more transparent interpretation for Eq. (3) which 
represents whole process of acquiring
a cloud of solute atoms by a mobile dislocation and eventual
immobilization. This 
can be seen  by defining $N_i = \int_{-\infty}^t K(t - t^{\prime}) N_m
(t^{\prime}) dt^{\prime}$, where the kernel $K(t)$ which represents
the fact that solute atoms are arriving at a certain rate has
been chosen to be $\alpha_m e^{-\alpha_i t}$. It can be easily
checked that Eq. (3) is the differential form of the definition of
$N_i$. One last remark about some of the terms appearing in these equations.
The first two
terms plus the last term in Eq.(1), and equivalent terms in
Eqs.(2), were derived starting from a velocity distribution
function of the dislocation segments [ Ananthakrishna \& Sahoo
1981a; Sahoo \& Ananthakrishna, 1982].   Thus, this model is 
an improved model to explain the 
instability in plastic flow. Equations (1-3) can be cast into 
a dimensionless form by using scaled variables:
\begin{equation}
x = N_m (\frac{\beta}{\gamma}), y = N_{im}(\frac{\beta}{\theta
V_m}), z = N_i (\frac{\beta\alpha_i}{\gamma\alpha_m})
,\tau=\theta V_m t,
\end{equation}
to get
\begin{eqnarray}
\dot{x} & = & (1-a)x -b_0x^2 -xy +y,
\\
\dot{y} & = & b_0\left(b_0x^2 -xy-y+az\right),
\\
\dot{z} & = & c(x-z),
\end{eqnarray}
where $a = \frac{\alpha_m}{\theta V_m}, b_0=\frac{\gamma}{\theta
V_m}, {\rm and }\, c=\frac{\alpha_i}{\theta V_m}$.  The dot
represents differentiation with respect to $\tau$. Equations (5-7)
are coupled set of nonlinear equations which support limit cycle
solutions for a range of parameters $a,b_0$ and $ c$, that are
physically relevant. $a$ refers to the concentration of the
solute atoms, $b_0$ refers to the reactivation of immobile
dislocations and $c$ to the time scales over which the slowing
down occurs. We have demonstrated the existence of limit
cycle solutions and also obtained approximate closed form
solutions for the limit cycles [Ananthakrishna \& Sahoo, 1981b;
Valsakumar \& Ananthakrishna, 1983].  In addition, the model has
been studied numerically.  Using the Orowan equation which
relates the rate of change of strain($\dot{S}$) to mobile dislocation
density and the mean velocity: $\dot{S}=bN_mV_m$, with $b$ as
the Burger's vector, steps on the creep curve follow
automatically since the densities of dislocations are
oscillatory.  Several experimental results are reproduced
[Ananthakrishna \& Sahoo, 1981b; Valsakumar \& Ananthakrishna,
1983].

\section{ Reductive Perturbative Approach}

For the sake of completeness, we outline the reductive
perturbative approach. 
Near the point of Hopf bifurcation of the system (Eqs.5-7),
corresponding to a critical value of the drive parameter, a pair
of complex conjugate eigenvalues and another real negative
eigenvalue exist for the linearized system of equations around
the steady state. As we approach the critical value from the
stable side, the real parts of the pair of complex conjugate
eigenvalues approach zero and hence the corresponding
eigenvectors get slower and slower. In contrast, the effect of
the change in the drive parameter on the real negative
eigenvalue is negligible.  Thus, {\it the eigenvector
corresponding to the real negative eigenvalue is the fast mode}.
For this reason, the slow modes are the ones which determine the
formation of new states of order.  {\it The reductive
perturbative method is a method where the asymptotic equation is
extracted in a systematic way} [Taniuti \& Wei, 1968; Newell \&
Whitehead, 1969; Kuramoto \& Tsuzuki, 1974; Mashiyama {\it et
al}., 1975; Richter {\it et al}., 1981].  The method involves in
first finding the eigenvectors corresponding to the slow modes
and looking for a solution in the subspace spanned by these
vectors. The effect of the nonlinearity is handled progressively
as a perturbation of the linear solution in power series of the
deviation from the critical value of the drive parameter.  It
may be worth emphasizing that this method is essentially the same
as reduction to center manifold. Indeed, the equivalence of the
center manifold theory [Carr, 1981; Guckenheimer \& Holmes, 1983; 
Troger \& Steindl, 1991] with the reductive perturbation has been
established [Chen {\it et al}, 1996]. Other techniques of
extracting amplitude equations have been devised whose end
results are basically the same.  For instance, perturbative
renormalization group method [Goldenfeld {\it et al} 1989; Chen
{\it et al}, 1996] and its recent extension on the basis of
envelope theory [Kunihiro, 1995 \& 1996] has also been developed
as a tool for global asymptotic analysis which can be used to
extract the amplitude equations.

In the present model, there is one simplification arising from 
the fact that $x$ can be identified as the fast (stable) mode, 
and hence this will be eliminated adiabatically. Then we derive 
the equation
for the slow dynamics through a power series expansion. Although
this adiabatic elimination is approximate, the result is found
to be in a reasonably good agreement with the accurate method
(which will be reported elsewhere) except for finer details.

Consider the Eqs.(5-7). It can be shown that there is a domain
of instability for the parameters $a$, $b_0$ and $c$.  The range of
values of $b_0$ is $\sim 0 - 10^{-2}$, that of $c$ is $\sim 0 - 10^{-1}$
while $a\sim 1$. Thus, $b_0 $ and $c $ $\ll 1$. Thus, there are 
three time scales that are in principle different from each other. 
Physically, they correspond to the mobile (fastest $a$), 
the Cottrell type ($c$) and the immobile (slowest $b_0$). 
Under these conditions we can eliminate one full equation
adiabatically. This aspect becomes transparent if the above
equations are written in terms of variables which are deviations
from the steady state. There is only one steady state defined
by:
\begin{equation}
x_{a} = z_{a} = {{1-2a+[(1-2a)^2+8b_0]^{1/2}}\over{4b_0}}, {\rm
and},y_{a}=1/2.
\end{equation}

\noindent
Defining new variables which are deviations from the steady
state
\begin{equation}
X=x-x_a,Y=y-y_a, {\rm and}{} \, Z=z-z_a,
\end{equation}
\noindent
Eqns. (5-7) take the form \begin{eqnarray} 
\dot{X} & = &-(\alpha X +\chi Y +b_0X^2 +XY ), \\
\dot{Y} & = & - b_0\left(\Gamma X+\delta Y-a Z -b_0 X^2 + XY \right),
\\
\dot{Z} & = & c(X-Z).
\end{eqnarray}
where
\begin{eqnarray}
\alpha = a +2b_{0}x_{a} +y_{a} -1, \chi = x_{a} -1,
\nonumber
\\
& &
\\ 
\Gamma = y_{a} -2b_{0}x_{a}, \delta = x_{a} +1.
\nonumber
\end{eqnarray}
\noindent
Now, we can rescale the time-like variable by
$\tau^{\prime}=b_0\tau$ and get
\begin{eqnarray}
b_0{{d{X}}\over{d\tau^{\prime}}} & = & -(\alpha X+\chi Y +b_0X^2
+XY ),
\\
{{d{Y}}\over{d\tau^{\prime}}} & = & - \left(\Gamma X +\delta Y
-a Z -b_0 X^2 + XY \right) ,
\\
{{d{Z}}\over{d\tau^{\prime}}} & = & {{c}\over{b_0}}(X-Z).
\end{eqnarray}
\noindent
It is clear that in the limit of small $b_0$, $\vert{X}\vert
\rightarrow \infty$,
unless the left hand side vanishes identically. Thus, we can
eliminate $X$ in favour of the other two variables and obtain:
\begin{eqnarray}
{{d{Y}}\over{d\tau}} & = & -b_{0}\left[a X+2(x_a + X)Y
-aZ\right], \\
\mbox{and}\hspace{.5cm}  {{d{Z}}\over{d\tau}} &= & c\left[X -Z\right],
\end{eqnarray}
\noindent
where
\begin{equation}
X = \frac{1}{2b_0}
\left(-(\alpha+Y)+[Y^2+2(\alpha-2b_{0}\chi)Y+\alpha^2]^{\frac{1}{2}} \right).
\end{equation}
\noindent
(The other root for $X$ is unphysical since it corresponds to
negative dislocation density.) It must be emphasized that this
adiabatic elimination becomes more exact as the value of the
parameter $b_0$ gets smaller. Equations (17-18) will be solved
reductive perturbatively. Writing these equations as a matrix
equation where the nonlinear part appears separately from the
linear part, we get
\begin{equation}
{{d\vec{R}}\over{d\tau}}= {\bf L}\vec{R} + \vec{N}
\label{mtrxeqn}
\end{equation}
\noindent
where
\begin{equation}
\vec{R} = \left( \begin{array}{c} Y\\ Z \end{array} \right), 
\end{equation}
\begin{equation}
{\bf L} = \left( \begin{array}{cc} c_{0} & ab_{0} \\
-\frac{c\chi}{\alpha} & -c \end{array} \right), \end{equation}
\noindent
with $c_{0} = b_0(\frac{a\chi}{\alpha} -2x_a)$ and the
nonlinear part, $\vec{N}$, is given by
\begin{equation}
\vec{N} = \left( \begin{array}{c}
-ab_0(X+\frac{\chi}{\alpha} Y) - 2b_{0}XY \\
c(X+\frac{\chi}{\alpha} Y) \end{array} \right)=\left(
\begin{array}{c}
\sum_{n=2}^{\infty}\xi_{n}Y^{n} \\
\sum_{n=2}^{\infty}\zeta_{n}Y^{n}
\end{array} \right).
\end{equation} 
\noindent
The coefficients $\xi_{n}$ and $\zeta_{n}$ appearing in the last
expression for $\vec{N}$ are functions of $ a$, $b_{0}$ and $c$
whose expressions for the required first few are given in
Eqs.(A8-A15) of Appendix A.

Consider stability of the fixed point as a function of the
parameter $c$. The eigenvalues, $\lambda_{\pm}$, of the matrix
${\bf L}$ are
\begin{equation}
\lambda_{\pm}=\frac{1}{2}\left(c_0 -c \pm \sqrt((c_0
-c)^2-8b_{0}x_a c)\right).
\end{equation}
\noindent
The fixed point becomes unstable when $c$ is less than $c_0$
($c$ is non-negative) and the discriminant of Eq.(24) is
negative giving a pair of complex conjugate eigenvalues. This
holds when the inequality $(3a-1)(1-2a^2)\geq 2b_{0}(2+a)^2$ is
satisfied. The instability region in the $\frac{1}{c}$ versus
$a$ plane for a fixed value of $b_{0}(=10^{-4})$ is shown in
Fig. 1.  It extends approximately between $a=\frac{1}{3}$ and
$a=\frac{1}{\sqrt{2}}$.

To get approximate analytical solution of Eq.(20), we follow the
reductive perturbative approach similar to that used by
Mashiyama {\it et al}. [1975] and Richter {\it et al}. [1981].
We choose $c=c_0 (1-\epsilon)$ with $0 < \epsilon \ll 1$ and
write the matrix ${\bf L}$ as a sum of two matrices, ${\bf L}=
{\bf L}_0 +\epsilon{\bf L}_1$, where ${\bf L}_0$ is the matrix
${\bf L}$ evaluated for $c=c_0$ and
\begin{equation}
{\bf L}_1 = \left( \begin{array}{cc} 0 & 0 \\ \frac{c_0
\chi}{\alpha}  & c_0
\end{array} \right).
\end{equation}
\noindent
The eigenvalues of ${\bf L}_0$ are $\lambda_{0
\pm}=\pm\imath\omega$,
where $\omega = \sqrt(2b_{0}c_0x_a)$.
Considering that the solution for $\vec{R}$ grows continuously out of the
critical eigenmodes, we can express it as
\begin{equation}
\vec{R}(\tau)=\Psi (\tau)e^{\imath\omega \tau}\vec{r_+} + \Psi^{*} (\tau)e^{-\imath\omega \tau}\vec{r_-},
\label{lincomb}
\end{equation}
\noindent
where $\Psi$ and $\Psi^*$ are time-dependent complex amplitudes while
$\vec{r}_\pm$ are right eigenvectors defined by 
${\bf L}_0\vec{r}_\pm = \pm\imath\omega\vec{r}_\pm$.
Similarly, we introduce left eigenvectors, $\vec{s}_\pm^T$
defined by $\vec{s}_\pm^T{\bf L}_0 = \pm\imath\omega\vec{s}_\pm^T$. 
(Note that $\vec{s}_+^T\vec{r}_- =\vec{s}_-^T\vec{r}_+= 0$. T refers to
transpose operation of the matrix concerned). Substituting this
expression for $\vec{R}$ in Eq.($\ref{mtrxeqn}$) and multiplying it by 
$\vec{s}_+^T$ from the left-hand side gives the equation
governing $\Psi$: 
\begin{equation}
e^{i\omega \tau} \frac {d \Psi} {d \tau} \,=\, \epsilon \mu \Psi e^{i\omega \tau}
\,-\, \epsilon \mu^*  \Psi^{*} e^{-i\omega \tau} \,+\, \sum^{\infty}_{n=2}  P_n 
(\psi e^{i\omega \tau} \,+\, c.c. )^n.
\label{amp1}
\end{equation}
\noindent
$\mu=\mu_1 + \imath\mu_2$ and $P_n$ are complex coefficients which are
functions of $a$, $b_0$ and $c$. Their expressions are,
respectively, given in Eqs. (A1) and (A4) in Appendix A. 
We express $\Psi$ in power series of $\epsilon^{\frac{1}{2}}$: 
\begin{equation}
\Psi \,=\, \epsilon^{\frac {1} {2}} \psi_1 \,+\, \epsilon \psi_2 \,+\, ...\\
\end{equation}
\noindent
and introduce slowly varying multiple time scales
$\tau_1(=\epsilon\tau)$, $\tau_2(=\epsilon^2\tau)$,... 
replacing $\frac{d}{d\tau}$ by
$\frac{\partial}{\partial\tau}+\epsilon\frac{\partial}{\partial\tau_1}
+\epsilon^2\frac{\partial}{\partial\tau_2}+...$. With this
substitution in Eq.($\ref{amp1}$), it can be solved successively
by equating terms of equal power in $\epsilon$ on both sides of
the equation. At first, $\cal{O}$($\epsilon^{\frac{1}{2}}$) terms 
give rise to the equation
\begin{equation}
\frac {\partial \psi_1} {\partial \tau} \,=\, 0\\
\end{equation}
\noindent
which means that, on the time scale of $\tau$, $\psi_1$ is constant. 
The next higher order, $\cal{O}(\epsilon)$, terms give the equation
\begin{equation}
\frac {\partial \psi_2} {\partial \tau} \,=\, P_2 (\psi_1 e^{i\omega \tau} \,+\,
c.c. )^2 e^{-i\omega \tau}\\
\end{equation}
\noindent
which can be integrated to get $\psi_2$.
$\cal{O}$($\epsilon^{\frac{3}{2}}$) terms give the equation
\begin{equation}
\frac{\partial \psi_3}{\partial \tau} + \frac{\partial
\psi_1}{\partial\tau_1} =  \mu \psi_1 - \mu^* \psi_1^* e^{-2i\omega \tau} 
+  \left[ 2P_2[(\psi_1 \psi_2e^{2i\omega\tau} 
+ \psi_1 \psi_2^*) + c.c.] + P_3 (\psi_1 e^{i\omega\tau}
+ c.c. )^3 \right] e^{-i\omega\tau},
\end{equation}
\noindent
from which we extract the slow dynamics:
\begin{equation}
\frac {\partial \psi_1} {\partial \tau_1} \,+\, \mu \psi_1 \,+\, \eta \vert
\psi_1 \vert^2 \psi_1
\label{slowdyn1}
\end{equation}
\noindent
Since to ${\cal O}(\epsilon^{\frac{1}{2}})$ $\Psi = \epsilon^{\frac{1}{2}}\psi_1$,
Eq.($\ref{slowdyn1}$) gives the dynamics for the complex amplitude $\Psi$:
\begin{equation}
\frac{d\Psi}{d\tau} = \epsilon\mu\Psi + \eta\vert\Psi\vert^{2}\Psi.
\label{cubic}
\end{equation}
\noindent   
The expression for the complex coefficient $\eta=\eta_1 +\imath\eta_2$
is given in Eq. (A2) in Appendix A.
Equation ($\ref{cubic}$) is a {\it cubic} TDGL equation where $\Psi$
is the complex amplitude for the slow mode which is also
referred to as complex order parameter (of the new state of
temporal order).  Its steady state 
solution gives the amplitude squared as
\begin{equation}
   \vert\Psi\vert^2= -\epsilon\frac{\mu_{1}}{\eta_{1}}
\label{amp}
\end{equation}
\noindent
and the associated  frequency,
 $\Omega$, (with $\Psi=\vert\Psi\vert e^{\imath\Omega\tau}$) as
\begin{equation}
   {\Omega}={\epsilon}\left(\mu_{2} -\frac{\eta_{2}}{\eta_{1}}\mu_{1}\right).
\label{freq}
\end{equation}
\noindent
Note that $\Psi$ describes the limit cycle oscillation.  
This solution exists provided $\eta_{1}$ is negative since $\mu_{1}$ 
is  positive (see expression of $\mu$ in Eq.(A1) in Appendix A).  
$\eta_1$ is found to be negative in a narrow region 
($0.4713\le a \le 0.5120$) as can be seen in Fig. 2.
In this case, since the amplitude of the slow modes grows continuously 
in proportion to $\epsilon^{\frac{1}{2}}$ (see Eq.($\ref{amp}$))
the transition is supercritical bifurcation
(continuous or `second order') 
 For other values of $a$ (within the instability range),
$\eta_{1}$
is positive implying that the transition is subcritical bifurcation 
(discontinuous or `first order'). 
One has to then go to quintic or even higher terms in the TDGL 
equation for obtaining an expression for the slow modes (order parameter). 
We have carried out the reductive 
perturbative method further to derive the  {\it quintic} TDGL equation: 
\begin{equation}
    \frac{d}{d\tau}\Psi=\epsilon\mu\Psi 
+\eta\vert\Psi\vert^{2}\Psi
  +\nu\vert\Psi\vert^{4}\Psi.
\label{quintic}
\end{equation}
\noindent
The steady state solution of this equation gives 
the amplitude squared as
\begin{equation}
     \vert\Psi\vert^2=\frac{1}{2}\left(-\frac{\eta_1}{\nu_1} 
+ \sqrt{(\frac{\eta_1}{\nu_1})^2
-4\epsilon\frac{\mu_1}{\nu_1}}\right),
\end{equation}
and the frequency as
\begin{equation}
   \Omega = \left(\eta_2  - \eta_1\frac{\nu_2}{\nu_1}\right)\vert\Psi\vert^2 + 
\epsilon\left(\mu_2 - \mu_1\frac{\nu_2}{\nu_1}\right).
\end{equation}
We found $\nu_{1}$ to be negative in the range $0.4538\le a\le0.5097$ 
(see Fig. 2). {\it This means a narrow region:
$0.4538\le a\le0.4713$ within the subcritical bifurcation exists over which 
the quintic TDGL equation has a solution} (see Fig. 2).

\section {Comparison with Experiments and Numerical Solutions}

\subsection{Comparison with Experiments}
To start with, consider the supercritical regime where
expressions for the amplitude and period of the limit cycle are simple. 
Using the steady state solution of the cubic TDGL equation, 
Eqs.($\ref{amp}$) \& ($\ref{freq}$), the dependence of 
 $\vert\Psi\vert^2$ and period, 
$P \sim \frac{1}{(\omega + \Omega)}$, of the limit cycle on $a$ are, to
the leading order, found to be $a^{-2}$ and $(const. + a^2)$,
respectively. The parameter $ a (=\alpha_m/\theta V_m)$ 
is a function of the applied stress ($\sigma_a$) and temperature
(T). As remarked earlier, $\alpha_m$ is proportional to the
concentration of the solute atoms, and therefore $\alpha_m \sim exp
(-E/kT)$. Using the standard expression [ Alexander 1986]
$V_m(\sigma_a, T) = V_0 (\sigma_a/\sigma_0)^m exp (- E_m/kT)$ (with $m>1$),
we get   
\begin{equation}
\frac{1}{a} \sim \left(\frac{\sigma_a}{\sigma_0}\right)^m e^{\frac{E-E_m}{kT}}.
\label{rel}
\end{equation}
From this we see that the amplitude of the limit cycle has a
strong increasing dependence on stress and an increasing
dependence on temperature if we
assume that $E > E_m$. In contrast, the period of the limit
cycle has a weak decreasing dependence on stress (as the leading
contribution is constant) and a decreasing
dependence on temperature.
Since stress and temperature are measurable quantities, 
our predictions can be compared with experimental results. 
The amplitude and period of the limit cycle are related,
respectively, to the amount of strain jumps and the period of the jumps 
on the creep curve through the Orowan equation. 
There are 
very few experiments in this mode of testing as mentioned 
in the introduction. The only experiment where this
dependence on stress and temperature has been measured is that by  
Zagoruyko {\it et al} [1977].
However, it is possible to translate the results from
experiments in constant strain rate case to the creep case 
and compare them with the results of this calculation.  
According to Zagoruyko {\it et al} [1977], the amplitude of 
the strain jumps increases with stress while its period 
has a weak but decreasing dependence on stress. Experiments from constant
strain rate case also exhibit the same trend when the results are 
translated in terms of constant stress experiments. It is well 
known that the amplitude of the stress drops decreases with 
applied strain rate. In fact, even the 
present model predicts this behaviour for the constant strain rate mode 
[Ananthakrishna \& Valsakumar 1982]. This implies
that the dependence of the amplitude of strain jumps
on stress should increase [ Kubin {\it et al} 1993]. 
[ This relation can 
be seen as follows. In constant strain rate case, the deformation rate is 
fixed and the stress developed in the sample is measured.
When the contribution to the plastic strain rate increases
due to increased dislocation motion (for whatever reasons), 
the stress has to fall in order to keep the applied strain rate constant. 
Thus, the relation between strain rate and stress is opposite.] 
Clearly, the general trend is consistent with the experimental results.
Fig. 3 shows the actual dependence of the amplitude and the period on $a$
in the domain of supercritical bifurcation (not the leading one as given
by the above expressions).
Zagarukuyko ${\it et al}$ [1977] also report that the amplitude
of the strain jumps increases while its period
decreases with temperature, which is  consistent with our
result ( provided $E > E_m $ ).

In the case of subcritical bifurcation, the dependence of the
amplitude of the strain jumps and the period on $a$ have similar behaviour
as in the case of the supercritical bifurcation. Eventhough, in this
case, leading order dependence on $a$ is difficult to obtain we
have  their actual dependence plotted in Fig. 4. It may be
noted that they have qualitatively the same behaviour as in 
Fig. 3 with the degree of dependence more pronounced in this case.

Experiments  in constant strain rate case show that the stress drops 
are seen to arise both abruptly as well as continuously  
[Kubin {\it et al}, 1988]. Translating this result to the constant 
stress case, it implies that the strain jumps can arise  both 
abruptly and continuously. This feature again follows from our calculations. 

\subsection{Comparison with Numerical Solutions}

Having derived the TDGL equation, we would first like to 
compare its result with the numerical solutions 
obtained via Eqs. (17-18) and via Eq.($\ref{mtrxeqn}$). Secondly, due to the 
fact that the quintic TDGL equation (Eq. $\ref{quintic}$) is 
valid in a limited domain, one suspects that higher order
nonlinearities are controlling the subcritical bifurcation,
which is quite unusual.
Therefore, we would like to investigate the numerical
 solution  
obtained by solving Eq.($\ref{mtrxeqn}$) keeping successive leading powers in
$Y$. This analysis confirms the above suspicion that higher order 
nonlinearities are important in 
this model outside the domain of validity of the quintic TDGL equation.

Using the steady state solution of the quintic TDGL equation, 
Eq.($\ref{quintic}$), in
Eq.($\ref{lincomb}$), we get an approximate expression for the limit cycle which is
usually called the secular motion. The equations governing the secular
motion are given by Eqs. (B5) and (B6) (hereafter called secular
equations) in Appendix B. This can be compared with the numerical solutions 
obtained using Eqs. (17-18) (hereafter called the reduced model) and with 
that obtained using Eq.($\ref{mtrxeqn}$) {\it where} the expansion in $Y$ of 
Eq. (23) is limited up to a certain power (hereafter called the 
$Y^n$-truncated model, n signifying the highest power of $Y$ after 
truncation). {\it At the outset we state that the reduced model has bounded
solution over the entire instability domain}.

In the case of the numerical solutions obtained from Eq.($\ref{mtrxeqn}$), 
we have found that while truncating the $Y$-expansion in Eq. (23) 
at $Y^3$ is sufficient to give a bounded solution for the range of 
supercritical bifurcation ($ 0.4713 < a < 0.512 $), it fails beyond this. 
On the other hand, truncating the $Y$-expansion at $Y^5$ does extend the 
domain of bounded solution ($ 0.4538 < a < 0. 512$) which matches with the
domain where the quintic TDGL equation works. 

Figure 5 shows the plots of the solutions to the secular equations 
(Eqs. B5 and B6), 
the reduced model (Eqs. 17-18) and the $Y^3$-truncated model 
(Eq. $\ref{mtrxeqn}$) for $a=0.5$ and $\epsilon=10^{-4}$. 
All of them are numerically almost indistinguishable.
We have verified that the solutions obtained by these three methods show
that the amplitude of the limit cycle scales with $\epsilon^{\frac{1}{2}}$.

Figure 6 shows the solutions obtained by the three ways for $a=0.468$
and $\epsilon=10^{-4}$ where, this time, the truncation is at $Y^5$.
The agreement between the three solutions is reasonably good. Part of the 
discrepancy between
the secular motion and the two numerical solutions can be attributed to
the range of values of the various terms 
($k$'s, $\ell$'s, $m$'s and $n$'s of Appendix C)
contributing to $\eta$ and $\nu$. (They range from $10^{-16}$ to $10^9$).
We have also verified that the amplitude of the limit cycle has a finite
jump for this case. 

For $a< 0.4538$, where the quintic TDGL equation is inadequate, 
$Y^5$-truncated model is
also inadequate. However, $Y^7$-truncated model has a bounded solution for 
the region $ 0.4385 < a < 0.5120$. In particular, 
Fig. 7 shows the plots of the solutions obtained by the $Y^7$-truncated 
model and that by the reduced model for $a=0.44$ and  $\epsilon=10^{-4}$.
{\it This implies that seventh order TDGL equation need to be considered
for this range of $a$}.  We have verified that $Y^9$ term extends the domain
of bounded solutions upto $ a = 0.4246 $. From this trend, it appears 
that very high powers in $Y$
need to be retained to cover the entire domain upto $a = 0. 333$. 

The region $a > 0.5120$ is even more interesting.
While the quintic TDGL equation is inadequate to give a  bounded solution, 
we found that retaining even upto the ninth power in $Y$ does not give a 
bounded solution for the entire range of values $0.512 < a < 0.707$ (even 
for those close to $a=0.512$). This suggests that this region is one 
where the nonlinearity is strong. On the other hand, as pointed out earlier
the reduced model
has limit cycle solution in this range as well. In particular, Fig. 8 
shows such a plot for $a=0.63$ and $\epsilon=10^{-4}$. 

Thus, the results of the TDGL equation, the $Y^n$-truncated model and
the reduced model are consistent with each other. Further, the numerical
solutions via the power series in $Y$ throws light
on the nature of nonlinearities governing the solution in 
different regions of instability.

\section{ Summary and Discussion}
We have carried out the reductive perturbative approach to the problem
of steps on creep curve and shown that the dynamics of the
system is described by a TDGL equation for the amplitude of the
slow modes (complex order parameter) $\Psi$ in the neighbourhood
of the bifurcation point. Since the above
derivation is valid only in the neighbourhood of the
critical value, the expression for 
$\Psi$ is valid only for small $\epsilon$. This has been exemplified by the
reasonable agreement between the secular motion and the numerical solution 
of the reduced model within the domain of the validity of the quintic
TDGL equation.

We have shown that both subcritical as well as supercritical bifurcations
are seen in the instability range of the parameter $a$. {\it While the 
subcritical bifurcation ( abrupt or `first order' transition) 
seen over most of the 
values of $a$ is consistent with our earlier calculation} 
[Ananthakrishna \& Sahoo, 1981b; Valsakumar \&  Ananthakrishna, 1983], 
{\it we see a supercritical bifurcation ( continuous or 
`second order' transition) over a 
narrow mid-range of} $a$. The existence of the supercritical bifurcation
was missed by us earlier. {\it The existence of both supercritical and
subcritical bifurcations is consistent with experiments.
In addition, the dependence of the amplitude and the periodicity 
of the strain jumps on stress and temperature are consistent
with experiments}. From this point of view, even though the 
derived TDGL equation is valid in 
a limited domain, the present calculation allows a direct mapping of 
these quantities. This calculation also helps us
to demonstrate the complicated dependence of the slow modes
on the original modes. This will serve as a warning to those
using hand waiving arguments for declaring certain modes as fast modes
and others as slow modes in modelling of such of these problems.

Looked at from the point of view of the properties of the model,
there are some interesting features. For the region outside the 
validity of the quintic TDGL equation, on the basis of a careful 
numerical analysis, we found that high order nonlinearities govern 
the nature of the subcritical bifurcation. It must be emphasized 
that this feature is unusual and this insight would not have 
been possible but for the numerical solutions obtained by
keeping successive higher powers in $Y$, which in itself 
is the basis for the derivation of the TDGL equation.

This unusual feature of the model interpreted in the language of
phase transitions give better insight and could also find applicability. 
For conventional models,  
the `free energy' is described by an expansion in power series 
of the order parameter. Even in the case of dynamic transition
such as the present one, it is possible to associate a 
`free energy' like function such that
$\partial \Psi\over\partial\tau$ $= -\frac{\delta F}{\delta \Psi^*}$, where 
$F[\Psi,\Psi^*]$ is the `free energy' like function. 
While the `free energy' for 
`second order' phase transitions is described by retaining up to fourth 
power in the order parameter, up to sixth power is conventionally required 
for  `first order' phase transitions (with the appropriate signs 
for the coefficients in the expansion). In the present case, however,
we need to go to as high as twelfth power (or more) of the order parameter 
to cover the interval $0.3333 < a < 0.4246$ (or equivalently twelfth power or 
higher in $Y$). In this case, including 
successively higher powers in $Y$  increases the domain of description.
On the other hand, in the 
interval of $0.512 < a < 0.707$, even retaining  upto tenth power
in $Y$ (which is the highest power we have checked) does not work. 
This must be contrasted with the existence of  bounded solution for 
the reduced model over the entire interval. {\it This feature is rather 
unusual and, to the best of our knowledge, we are not aware of any 
other model which exhibits this property}.

\subsection*{ Acknowledgement}
One of us (M.B.) would like to thank the International Program in Physical
Sciences, Uppsala University (Sweden) for offering a fellowship to study
at Indian Institute of Science. Part of this work has been supported
by IFCPAR Grant No. 1108-1.

\newpage

\noindent
\begin{description}
\item {\bf References}
\item Aifantis, E. C. [1988] ``On the problem of dislocation
           patterning and persistent slip bands'' in 
           {\it Non Linear Phenomena in Materials Science I}, eds. 
           Kubin, L. P. \& Martin, G. (Trans Tech, Switzerland), pp. 397-406.
\item Alexander, H. [1986] ``Dislocations in covalent crystals''
in {\it Dislocations in Solids} ed. Nabarro, F. R. N. (North-Holland), 151.
\item Ananthakrishna, G. [1993] ``Formation, propagation of
          bands and chaos in jerky flow'',
          {\it Scripta. Metall.} {\bf 29}, 1183-1188.
\item Ananthakrishna, G. \& John, T. M. [1990] ``Instabilities and chaos in 
          plastic flow'', in {\it Directions in Chaos} Vol. 3, ed. Hao Bai-lin
          (World Scientific, Singapore), pp. 133-148.
\item Ananthakrishna, G., Fressengeas C., Grosbras, M., Vergnol, J.,
           Engelke, C., Plessing, J., Neuhaeuser, H., Bouchaud, E., Planes, J. 
           \& Kubin, L. P. [1995a] ``On the existence of chaos in jerky flow,'' 
           {\it Scripta Metall.} {\bf 32}, 1731-1737.
\item Ananthakrishna, G., Kubin, L. P. \& Martin, G. (eds.) [1995b]
           {\it Solid State Phenomena}  Vol. 42\&43 (Scitec, Switzerland).
\item Ananthakrishna, G. \& Noronha, S. J. [1995] ``Chaos in jerky
           flow: Theory and experiment'' in {\it Solid State Phenomena}
           Vol. 42\&43, eds. Ananthakrishna, G., 
           Kubin, L. P. \& Martin, G. (Scitec, Switzerland) , pp. 277-286.
\item Ananthakrishna G. \& Sahoo, D. [1981a] ``A statistical
           dislocation dynamics with application to LiF'', 
           {\it J. Phys. D} {\bf 14}), 699-713.
\item Ananthakrishna, G. \& Sahoo, D. [1981b] ``A model based on 
          nonlinear oscillations to explain jumps on creep curves'',
          {\it J. Phys. D} {\bf 14}, 2081-2090. 
\item Ananthakrishna, G. \& Valsakumar, M. C. [1982] ``Repeated
          yield drop phenomena: a temporal dissipative structure'', 
          {\it J. Phys. D} {\bf 15}, L171-L175.
\item Ananthakrishna, G. \& Valsakumar, M. C. [1983] ``Chaotic flow in a
          model for repeated yielding'', {\it Phys. Lett. A} {\bf 95}, 69-71.
\item Ananthakrishna, G., Fressengeas C., Grosbras, M., Vergnol, J.,
           Engelke, C., Plessing, J., Neuhaeuser, H., Bouchaud, E., Planes, J. 
           \& Kubin, L. P. [1995] ``On the existence of chaos in jerky flow,'' 
           {\it Scripta Metall.} {\bf 32}, 1731-1737.
\item Bodner, S. R. \& Rosen, A. [1967] ``Discontinuous yielding of
          commercially-pure aluminium'', {\it J. Mech. Phys. Solids} 
          {\bf 15}, 63-77.
\item Brindley, B. J. \& Worthington, P. J. [1970] ``Yield-point
          phenomena in substitutional alloys'', {\it Metall. Reviews} 
          {\bf 145}, 101-114.
\item Carr, J. [1981] {\it Applications of Center Manifold Theory}
          (Springer-Verlag, Berlin).
\item Chen, L. Y., Goldenfeld, N. \& Oono, Y., [1996] ``The
          Renormalization Group and Singular Perturbations: Multiple-
          Scales, Boundary Layers and Reductive Perturbation Theory'',
          {\it Phys. Rev. E} {\bf 54}, 376-394.
\item Chihab, K., Estrin, Y., Kubin, L. P. \& Vergnol, J. [1987]
          ``The kinetics of the Portevin-Le Chatelier effect in an 
          Al-5 at\% Mg alloy'', {\it Scripta. Metall.} {\bf 21}, 203-208.
\item Cottrell, A. H. [1953] ``A note on the Portevin-Le Chatelier
          effect'', {\it Phil. Mag.} {\bf 44}, 829-832.
\item Da Silveira, T. L. \& S. N. Monteiro, S. N. [1979] ``Jumps
          in the creep curve of austentic stainless steels'', 
          {\it Met. Trans. A} {\bf 10}, 1795-1796.
\item Estrin, Y. \& Kubin, L. P. [1989] ``Collective dislocation 
          behavior in dilute alloys and the Portevin-Le Chatelier effect'',
          {\it J. Mech. Behaviour of Materials} {\bf 2}, 255-292.
\item Goldenfeld, N., Martin, O. \& Oono, Y. [1989]
``Intermediate Asymptotics and Renormalization Group Theory'', {\it J. Sci.
          Comp.} {\bf 4}, 355.
\item Guckenheimer, J. \& Holmes, P. [1983] {\it Nonlinear Oscillations, 
          Dynamical Systems, and Bifurcations of Vector Fields} vol. 42
of {\it Applied Math. Sciences} (Springer-Verlag, Berlin), 123-145. 
\item Hahner, P. \& Kubin, L. P. [1992] ``Coherent Propagative Structures
           in Plastic Deformation: A Theory of Ludlers Bands in Polycrystals''
           in {\it Solid State Phenomena} Vol. 3\&4, eds. 
           Kubin, L. P. \& Martin, G. (Trans Tech, Switzerland), pp. 385-402. 
\item Hahner, P. [1993] ``Modelling the spatiotemporal aspects of the 
          Portevin-Le Chatelier effect'', {\it Mater. Sci. and Eng.}
          {\bf 164}, 23-34.
\item Hall, E. O. [1970] {\it Yield Point Phenomena in Metals and
          Alloys} (Macmillan, London).
\item Kubin, L. P., Chihab, K. \& Estrin, Y. [1988] ``The rate
dependence of the Portevin-Le Chatelier effect'', 
{\it Acta Metall.} {\bf 36}, 2707-2718.
\item Kubin, L. P. \& Estrin, Y. [1990] ``Evolution of dislocation 
         densities and the critical conditions for the Portevin-Le Chatelier
         effect'', {\it Acta Metall.} {\bf 38}, 697-708.
\item Kubin, L. P. \& Estrin, Y. (organizers) [1993] ``Viewpoint 
          Set No. 21: Propagative instabilities'', {Scripta Metall.} 
          {\bf 29}, 1147-1188.
\item Kubin, L. P. \& Lepinoux, J. [1988] ``The dynamic Organization of
dislocation structures'', {\it Strength of Metals and Alloys} (ICSMA 8), 
vol. 1, eds. Kettunen, P. O., Lepisto, T. K., \& Lehtonen, M. E., 
(Pergamon Press ,Oxford), pp. 35-59.
\item Kubin, L. P. \& Martin, G. (eds.) [1988]  {\it Solid State Phenomena}
           Vol. 3\&4 (Trans Tech, Switzerland).
\item Kunihiro, T. [1995] ``A Geometrical Formulation of the
Renormalization Group Method for Global Analysis'', 
{\it Prog. Theor. Phys.} {\bf 94} 503-514. 
\item Kunihiro, T. [1996] ``The Renormalization Group Method
Applied to Asymptotic Analysis of Vector Fields'', in hep-th/9609045.
\item Kuramoto, Y. \& Tsuzuki, T. [1974] ``Reductive perturbative
           approach to chemical instabilities'', {\it Prog. Theor. Phys.} 
           {\bf 52}, 1399-1401.
\item Lubahn, J. D. \& Felgar, R. P. [1961] 
          {\it Plasticity and Creep of Metals} (John Wiley, New York).
\item Martin, G. \& Kubin, L. P. (eds.) [1992] {\it Solid State Phenomena}
           Vol.23\&24 (Trans Tech, Switzerland).
\item Mashiyama, H., Ito, A. \& Ohta, T. ``Fluctuations and phase
           transitions far from equilibrium'', {\it Prog. Theor. Phys.} 
           {\bf 54}, 1050-1066.
\item Newell, A. C. \& Whitehead, J. A. [1968] ``Finite 
           bandwidth, finite amplitude convection'', 
           {\it J. Fluid Mech.} {\bf 38}, 279-303.
\item Noronha, S. J., Ananthakrishna, G., Quaouire, L. \& Fressengeas, C.
           [1997] ``Chaos in Jerky Flow- Experimental verification of a
           theoretical prediction'', {\it Pramana}, in press.
\item Noronha, S. J., Ananthakrishna, G., Quaouire, L., Fressengeas, C.
           \& Kubin, L. P. [1996] ``Chaos in the Portevin-Le Chatelier Effect'',
            submitted to Int. J. Bif. \& Chaos.
\item Penning, P. [1972] ``Mathematics of the Portevin-Le Chatelier 
          effect'', {\it Acta Metall.} {\bf 20}, 1169-1175.
\item Quaouire, L. \& Fressengeas, C. [1995] ``Dynamical analysis of the
           Portevin-Le Chatelier effect'' in {\it Solid State Phenomena}
           Vol. 3\&4 (Trans Tech, Switzerland), eds. Ananthakrishna, G., 
           Kubin, L. P. \& Martin, G. (Scitec, Switzerland), pp. 293-302.
\item Richter, P. H., Procaccia, I. \& Ross, J. [1981] ``Chemical 
           Instabilities'' in {\it Advances in Chemical Physics} Vol. 43, 
           eds. Prigogine, I, \& Rice, S. (Wiley-Interscience, New York), 
           pp. 217-268.
\item Sahoo, D. \& Ananthakrishna, G.  [1982] ``A phenomenological
           dislocation transformation model for the mobile fraction in simple
           systems'', {\it J. Phys. D} {\bf 15}, 1439-1449.
\item Stejskalova, V., Hammersky, M., Luckac, P., Vostry, P. \&
Spursil, B. [1981] {\it Czech. J. Phys.} {\bf B31}, 195.
\item Taniuti, T. \& Wei, C. C. [1968] ``Reductive perturbative 
           method in nonlinear wave propagation I'', 
           {\it J. Phys. Soc. Japan} {\bf 24}, 941-946.
\item Troger, H. \& Steindl, A. [1991] {\it Nonlinear Stability
and Bifurcation Theory} (Springer-Verlag, Wien, New York), pp. 52-68.   
\item Valsakumar, M. C. \& Ananthakrishna, G. [1983] ``A model based on 
          nonlinear oscillations to explain jumps on creep curves: II.
          Approximate solutions'', {\it J. Phys. D} {\bf 16}, 1055-1068.
\item Venkadesan, S., Murthy, K. P. N., \& Valsakumar, M. C. [1995]
`` Chaotic Flow in Al-Mg Alloy'' in {\it Solid State Phenomena}
           Vol. 42\&43, eds. Ananthakrishna, G., 
           Kubin, L. P. \& Martin, G. (Scitec, Switzerland), pp. 287-292.
\item Zagoruyko, L. N., Osetskiy A. I. \& Soldatov, V. P. [1977] 
          ``Discontinuous deformation of zinc single crystals under creep
          conditions'', {\it Phys. Met. Metallogr.}  {\bf 43}(5), 156-164.

\end{description}

\newpage

\section*{Appendix A}

In this appendix we give the expressions for the coefficients appearing
in the TDGL equation up to the quintic term, i.e. the expressions for
$\mu$, $\eta$ and $\nu$ in the equation 
$\frac{d}{d\tau}\Psi=\epsilon\mu\Psi 
+\eta\vert\Psi\vert^{2}\Psi
  +\nu\vert\Psi\vert^{4}\Psi$.
\noindent
They are given by
\setcounter{equation}{0}
\renewcommand{\theequation}{A\arabic{equation}}
\begin{equation}
\mu\equiv \frac{1}{2}(c_0-\imath\omega),
\end{equation}
\begin{equation}
\eta\equiv 2P_2(k_1+k_2+k_2^*+k_3^*) + 3P_3,
\end{equation}
\noindent
and
\begin{equation}
\nu\equiv 2P_2Q_2 +3P_3Q_3 +4P_4Q_4 + 10P_5,
\end{equation}
\noindent
where
\begin{equation}
P_n\equiv\frac{(ab_0)^{n-1}}{2\omega}\left[\omega\xi_n -\imath(c_0\xi_n +ab_0\zeta_n)\right],
\end{equation}
\noindent
and
\begin{equation}
Q_2\equiv(k_1+k_3^*)\ell_3 +(k_2+k_2^*)\ell_3^*+(k_1^*+k_3)(\ell_2+\ell_4^*) + m_5+m_6+m_6^*+m_7^*,\\
\end{equation}
\begin{equation}
Q_3\equiv 2(k_1+k_3^*)(k_1^*+k_2+k_2^*+k_3)+(k_2+k_2^*)^2+\ell_2+\ell_3+2\ell_3^*+\ell_4^*,\\
\end{equation}
\begin{equation}
Q_4\equiv k_1^*+k_3+3(k_1+k_2+k_2^*+k_3^*).
\end{equation}
The expressions for $\xi_n$'s and $\zeta_n$'s used in Eq.(A5) are given by
\begin{equation}
\xi_2\equiv\frac{b_0\chi}{\alpha^3}\left[2\alpha^2 -a(\alpha-b_0\chi)\right],
\end{equation}
\begin{equation}
\xi_3\equiv\frac{b_0\chi}{\alpha^5}(\alpha - b_0\chi)(a\Delta - 2\alpha^2),
\end{equation}
\begin{equation}
\xi_4\equiv\frac{\alpha^2 - \Delta^2}{16\alpha^7}\left[a(\alpha^2 - 5\Delta^2) + 8\Delta\alpha^2\right],
\end{equation}
\begin{equation}
\xi_5\equiv -\frac{\alpha^2-\Delta^2}{16\alpha^9}\left[a\Delta(3\alpha^2 - 7\Delta^2) - 2\alpha^2(\alpha^2 - \Delta^2)\right],
\end{equation}
\noindent
and
\begin{equation}
\zeta_2\equiv\frac{c_0}{4b_0\alpha^3}(\alpha^2 - \Delta^2),
\end{equation}
\begin{equation}
\zeta_3\equiv -\frac{\Delta}{\alpha^2}\zeta_2,
\end{equation}
\begin{equation}
\zeta_4\equiv - \frac{\alpha^2 - 5\Delta^2}{4\alpha^4}\zeta_2,
\end{equation}
\begin{equation}
\zeta_5\equiv\frac{\Delta(3\alpha^2 - 7\Delta^2)}{4\alpha^6}\zeta_2,
\end{equation}
\noindent
where
\begin{equation} 
\Delta\equiv \alpha- 2b_0\chi.
\end{equation}  

Expressions for $k$'s, $\ell$'s, $m$'s and $n$'s are given in Appendix C.

\section*{Appendix B}

In this appendix the equations describing the secular motion are derived 
for the case when the order parameter, $\Psi$, is found from the quintic 
TDGL equation (Eq. 44). 

From Eq. (39), 
\setcounter{equation}{0}
\renewcommand{\theequation}{B\arabic{equation}}
\begin{equation}
Y(\tau)=ab_0(\Psi e^{\imath\omega\tau} + c. c.),
\end{equation} 
and
\begin{equation}
Z(\tau)=(-c_0+\imath\omega)\Psi e^{\imath\omega\tau} + c. c.,
\end{equation}
where the complex amplitude is
$\Psi=\epsilon^{\frac{1}{2}}\psi_1 +\epsilon\psi_2 +...+\epsilon^{\frac{5}{2}}\psi_5$
with  $\psi_2,..., \psi_5$ given in terms of functions of $\psi_1$ and phase
factors (see Appendix C). On the other hand, to 
${\cal O}(\epsilon^{\frac{1}{2}})$,
\begin{equation}
\Psi=\epsilon^{\frac{1}{2}}\psi_1,
\end{equation}
which, upon inverting, gives $\psi_1$ in terms of $\Psi$:
\begin{equation}
\psi_1=\epsilon^{-\frac{1}{2}}\vert\Psi\vert e^{\imath\Omega\tau}.
\end{equation}
\noindent
Substituting the $\psi_2,...,\psi_5$-expressions from Appendix C in Eqs.
(B1) and (B2) and, then, using Eq. (B4)  leads to obtaining the secular 
equations given by:
\begin{equation}
Y=2ab_0 \,P_+,
\end{equation}
\noindent
and
\begin{equation}
Z=-2c_0 \,P_+ \,-2\omega \,P_-.
\end{equation}
\noindent
where
\begin{eqnarray}
P_{\pm} &\equiv &|  \Psi | \, F_{\pm}(\Omega_c\,\tau) + K_{13}^{\pm}\, |\Psi|^{2} 
     F_{\pm}(2\Omega_{c} \, \tau + \theta_{k_{13}}^{\pm}) + |\Psi|^{2} \, 
      G_{\pm}(k_2) \nonumber\\
  &   &  +\epsilon\, L_1 \, |\Psi|\, F_{\pm} \left(\Omega_c\, \tau - 
         \theta_{\ell_{1}}\right)
        +  L_{24}^{\pm} \, |\Psi|^{3}\, F_{\pm} \left(3\Omega_c\, \tau +
          \theta_{\ell_{24}}^{\pm}\right)
        + L_3 \, |\Psi|^{3}\, F_{\pm} \left(\Omega_c\, \tau - 
        \theta_{\ell_{3}}\right)   \nonumber\\
  &  &  + \epsilon \, M_{13}^{\pm} \, |\Psi|^{2}\, F_{\pm} \left(2\Omega_c\,
         \tau + \theta_{m_{13}}^{\pm} \right)
        + \epsilon \, |\Psi|^2 \, G_{\pm} (m_{2}) +
          M_{48}^{\pm} \, |\psi|^{4} \, F_{\pm} \left(4\Omega_c\, \tau 
        + \theta_{m_{48}}^{\pm} \right)   \nonumber\\
   &  &  +  M_{57}^{\pm} \, |\Psi|^{4} \, F_{\pm} \left(2\Omega_c\, \tau + 
            \theta_{m_{57}}^{\pm} \right)
          + |\psi|^{4} \, G_{\pm} (m_{6})  +  \epsilon^{2}\, N_1 \, |\Psi| \, F_{\pm} \left(\Omega_c\, \tau
             - \theta_{n_{1}}\right)  \nonumber\\
   &  &  +\epsilon\, N_{24}^{\pm} \, |\Psi|^{3} \, F_{\pm} 
           \left(3\Omega_c\, \tau + \theta_{n_{24}}^{\pm}\right)
          +\epsilon\, N_3 \, |\Psi|^{3} \, F_{\pm} \left(\Omega_c\, \tau 
            - \theta_{n_{3}}\right) +  N_{59}^{\pm} \, |\Psi|^{5} \, F_{\pm} \left(5\Omega_c\, \tau 
               + \theta_{n_{59}}^{\pm}\right)  \nonumber\\
   &  &  + N_{68}^{\pm} \, |\Psi|^{5}\, F_{\pm} \left(3\Omega_c\, \tau 
         + \theta_{n_{68}}^{\pm}\right)  
         + N_{7}^{\pm} \, |\Psi|^{5}\, F_{\pm} \left(\Omega_c\, \tau 
          - \theta_{n_{7}}^{\pm}\right),      
\end{eqnarray}

with

\begin{eqnarray}
F_{+} (\phi) & \equiv & cos(\phi), \\
F_{-} (\phi) & \equiv & sin(\phi), \\
G_{+}(u) & \equiv &  Real (u),\\
G_{-}(u) & \equiv & Im (u) ,\\
\Omega_{c} & \equiv & \omega + \Omega,\\
V_{ij}^{\pm} & \equiv & | v_{i}  \pm  v^{\star}_{j}|,\\
V_{i} & \equiv & | v_{i}|,\\
\Theta^{\pm}_{v_{ij}} & \equiv & \sin^{-1} \left(\frac{Real(v_{i} \pm v_{j}^{\star})}
{|v_{i} \pm v_{j}^{\star}|}\right) ,
\end{eqnarray}
\noindent
where $V$ stands for either $K$, $L$, $M$ or $N$, while $v$ stands for either
$k$, $\ell$, $m$, or $n$ correspondingly. Expressions for $K$, $L$, $M$ 
and for $N$ can be obtained from $k$'s, $\ell$'s, $m$'s and $n$'s which 
are given in Appendix C.

\section*{Appendix C}

Expressions for $\psi_2$, $\psi_3$, $\psi_4$, $\psi_5$ (with a factor
$e^{\imath\omega\tau}$) are  
\setcounter{equation}{0}
\renewcommand{\theequation}{C\arabic{equation}}
\begin{eqnarray}
\psi_{2}\, e^{i\omega \tau} & = & k_{1}\, \psi_{1}^{2}\, e^{2i\omega
     \tau} + k_{2} |
     \psi_{1} |^{2} + k_{3}\, \psi_{1}^{\star^{2}}\, e^{-2i\omega \tau},\\
\psi_{3}\, e^{i\omega \tau} & = & \ell_{1}\, \psi_{1}^{\star}\,
     e^{-i\omega \tau}+\ell_{2}\, \psi_{1}^{3} \, e^{3i\omega \tau}+
     \ell_{3}|\psi_{1}|^{2}\, 
     \psi_{1}^{\star}\, e^{-i\omega \tau} + \ell_{4}\, \psi_{1}^{\star^{3}}\,
      e^{-3i\omega \tau},\\
\psi_{4}\, e^{i\omega \tau} & = & m_{1}\, \psi_{1}^{2} \, e^{2i\omega \tau}
     + m_{2}\,| 
     \psi_{1}^{2}| + m_{3} \, \psi_{1}^{\star^{2}}\, e^{-2i\omega \tau} +
      m_{4} \, \psi_{1}^{4}\, e^{4i\omega \tau}+ m_{5} \, |\psi_{1}|^{2}\,
      \psi_{1}^{2}\, e^{2i\omega \tau}\nonumber \\ 
 &     &+ m_{6} \, |\psi_{1}|^{4} +  m_{7} \, |\psi_{1}|^{2}\,
     \psi_{1}^{\star^{2}}\, e^{-2i\omega \tau}+ m_{8} \,
      \psi_{1}^{\star^{4}}\, e^{-4i\omega \tau} , \\
      \psi_{5}\, e^{2i\omega \tau} & = & n_{1} \,\psi_{1}^{\star}\, 
      e^{-i\omega  \tau} + n_{2}\,\psi_{1}^{3}\, e^{3 i \omega \tau} + n_{3}\,
      |\psi_{1}|^{2}\psi_{1}^{\star} \, e^{-i\omega \tau} + n_{4}
      \,\psi_{1}^{\star^{3}}\, e^{-3i \omega \tau} + n_{5} \, \psi_{1}^{5}\,
      e^{5i \omega \tau}\nonumber \\ 
 &     & + n_{6}|\psi_{1}|^{2} \psi_{1}^{3}\, e^{3i \omega\tau} 
+ n_{7}\, |\psi_{1}|^{4}\,  \psi_{1}^{\star}\, e^{-i\omega \tau} 
      + n_{8} \,|\psi_{1}|^{2}\, \psi_{1}^{\star^{3}}\, e^{-3i\omega \tau}+
      n_{9}\, \psi_{1}^{\star^{5}}\, e^{-5 i \omega \tau},
\end{eqnarray}
\noindent
where
\begin{eqnarray}
k_{1}  & \equiv & - \frac{i}{\omega}\, P_{2},\\
k_{2}  & \equiv &  \frac{2i}{\omega}\, P_{2},\\
k_{3}  & \equiv &  \frac{i}{3\omega}\, P_{2},\\
\ell_{1}  & \equiv & - \frac{i}{2\omega}\, \mu^{\star},\\
\ell_{2} & \equiv & -
         \frac{i}{2\omega}\,\left[2P_{2}(k_{1}+k_{3}^{\star})+P_{3}\right],\\
\ell_{3} & \equiv & 
      \frac{i}{2\omega}\,\left[2P_{2}(k_{1}^{\star}+k_{2}^{\star}+k_{2} +
      k_{3})+3P_{3}\right],\\
\ell_{4} & \equiv & 
      \frac{i}{4\omega}\,\left[2P_{2}(k_{1}^{\star}+k_{3})+P_{3}\right],\\
m_{1} & \equiv & -\frac{i}{\omega}\left[-\mu\, k_{1} - \mu^{\star}\,
      k_{3}^{\star} + 2 P_{2} \, \ell_{1}^{\star} \right],\\
m_{2} & \equiv & \frac{i}{\omega}\left[-\mu^{\star}\,(k_{2}+ k_{2}^{\star})+
      2 P_{2} (\ell_{1}+\ell_{1}^{\star}) \right],\\
m_{3} & \equiv & \frac{i}{3\omega}\left[\mu\, k_{3} -
      \mu^{\star}(k_{1}^\star+ 2k_{3}) +2P_2 \ell_{1}\right],\\
m_{4} & \equiv & -
      \frac{i}{3\omega} \left\{P_{2}\left[(k_{1}+k_{3}^{\star})^{2}+ 2
      (\ell_{2}+\ell_{4}^{\star})\right]+ 3P_{3} (k_{1}+k_{3}^{\star}) + P_{4}
      \right\},\\
m_{5} & \equiv & -
      \frac{i}{\omega}\left\{2P_{2}\left[(k_{1}+k_{3}^{\star})(k_{2}+k_{2}^{\star})
      +\ell_{2}+\ell_{3}^{\star} +\ell_{4}^{\star}\right]+ 3P_{3}
      \left[2(k_{1}+k_{3}^{\star}) +k_{2} +k_{2}^{\star}\right]\right. \nonumber \\
 &  &     + 4P_{4}- 2k_{1}\,\eta  \left. \right\},\\
m_{6} & \equiv &  \frac{i}{\omega}\left\{P_{2}\left[2(k_{1}+k_{3}^{\star})
                     (k_{1}^\star+k_{3})+(k_{2}+k_{2}^{\star})^{2} 
                     +2(\ell_{3} +\ell_{3}^{\star})\right]  \right. \nonumber\\
&  &      + 3P_{3}\left[k_{1}+k_{1}^{\star} + 2(k_{2}
                +k_{2}^{\star})+k_{3}+k_{3}^{\star}\right]
                + 6P_{4}-2k_{2}\eta_1  \left. \right\},\\
m_{7} & \equiv & 
       \frac{i}{3\omega}\left\{2P_{2}\left[(k_{1}^{\star}+k_{3})(k_{2}^{\star}+k_{2}) 
       +\ell_{2}^{\star}+\ell_{3} +\ell_{4}\right]+ 3P_{3}
       \left[2(k_{1}^{\star}+k_{3}) + k_{2}
       +k_{2}^{\star}\right]\right. \nonumber \\
 & &     + 4P_{4}-2k_{3} \eta^{\star}\left. \right\},\\
m_{8} & \equiv & 
       \frac{i}{5\omega}\left\{P_{2}\left[(k_{1}^{\star}+k_{3})^{2}
       +2(\ell_{2}^{\star} +\ell_{4})\right]
       + 3P_{3}(k_{1}^{\star} + k_{3})+P_{4} \right\},\\
n_{1} & \equiv & \frac{i}{2 \omega}\left(\mu - \mu^{\star}\right)
       \ell_{1},\\
n_{2} & \equiv & -\frac{i}{2 \omega} \left\{ 2 P_{2}\left[\ell_{1}^{\star} (k_{1}+ k_{3}^{\star}) +
       m_{1} + m_{3}^{\star}\right]  +3P_{3}\, \ell_{1}^{\star} -2\mu\ell_{2} - \mu^{\star}
       \ell^{\star}_{4} \right\},\\
n_{3} & \equiv & \frac{i}{2\omega} \left\{ 2 P_{2}\left[\ell_{1}(k_{2}+ k_{2}^{\star} ) +
       \ell^{\star}_{1}(k_{1}^{\star} + k_{3}) + m_{1}^{\star} +m_{2}^{\star}
       +m_{2}+ m_{3}\right]\right. \nonumber\\
&  &   + 3P_{3}(\ell_{1}^{\star}+2\ell_{1})\left.-\eta^{\star} \, \ell_{1} - 2
       \mu^{\star}\,\ell_{3} \right\},\\
n_{4} & \equiv & \frac{i}{4\omega} \left\{ 2 P_{2}\left[\ell_{1}(k_{1}^{\star}+ k_{3}) +
      m_{1}^{\star} + m_{3}\right] + 3P_{3}\, \ell_{1}-\mu^{\star} \ell_{2}^{\star} +
       (\mu - 3\mu^{\star})\ell_{4} \right\},\\
n_{5} & \equiv & -
       \frac{i}{4\omega}\left\{2P_{2}\left[m_{4}+m_{8}^{\star} +
       (k_{1}+k_{3}^{\star})
       (\ell_{2}+\ell_{4}^{\star})\right] + 3P_{3}
       \left[(k_{1}+k_{3}^{\star})^2 +\ell_{2} + \ell_{4}^{\star}\right] \right.\nonumber \\
&      &  +4P_{4}(k_{1}+k_{3}^{\star}) +P_{5}\left. \right\},\\
n_{6} & \equiv & -\frac{i}{2\omega}\left\{ 2P_{2}\left[m_{4} +
      m_{5} +m_{7}^{\star} +
       m_{8}^{\star} + \ell_{3}^{\star}(k_{1}+k_{3}^{\star}) + (\ell_{2}
      +\ell_{4}^{\star}) (k_{2} + k_{2}^{\star}) \right] \right.  \nonumber \\
 &   & + 3P_{3}
       \left[\ell_{3}^{\star} + 2(\ell_{2} + \ell_{4}^{\star}) + (k_{1} +
        k_{3}^{\star})^{2} + 2(k_{1}+k_{3}^{\star})(k_{2}+k_{2}^{\star}) \right]\nonumber \\
 &  &  + 4P_{4} \left[ k_{2} + k_{2}^{\star} + 3(k_{1}+k_{3}^{\star}) \right]
       + 5 P_{5}\left.-3\eta \ell_{2} \right\},\\
n_{7} & \equiv & \frac{i}{2\omega}\left\{ 2P_{2}\left[m_{5}^{\star}  +m_{6}^{\star}+
       m_{6} +m_{7} + (\ell_{2}^{\star}+\ell_{4})(k_{1}+k_{3}^{\star}) + \ell_{3}
       (k_{2} + k_{2}^{\star})+\ell_{3}^{\star}(k_{1}^{\star}+k_{3}) \right] \right. \nonumber\\
&  & + 3P_{3} \left[ (k_{2} + k_{2}^{\star})^{2} +
       2(k_{1}^{\star}+k_{3})(k_{1}+k_{2}+k_{2}^{\star}+k_{3}^{\star})+\ell_{2}^{\star}
       + \ell_{3}^{\star} + 2\ell_{3}+\ell_{4}\right] \nonumber\\
&   & + 4P_{4} \left[ k_{1} + k_{3}^{\star} + 3(k_{1}^{\star}+k_{2}^{\star}+k_{2}
       +k_{3})\right] +10 P_{5}\left.-(\eta
        +2\eta^{\star})\ell_{3} \right\},\\
n_{8} & \equiv & \frac{i}{4\omega}\left\{2P_{2}\left[m_{4}^{\star} + m_{5}^{\star} +m_{7} +
       m_{8} + \ell_{3}(k_{1}^{\star}+k_{3}) + (\ell_{2}^{\star}
       +\ell_{4}) (k_{2} + k_{2}^{\star}) \right] \right. \nonumber \\
&  &  + 3P_{3}\left[  (k_{1}^{\star} +
      k_{3})^{2} + 2(k_{2}+k_{2}^{\star})(k_{1}^\star+k_{3})+2 (\ell_{2}^{\star} +
      \ell_{4}) + \ell_{3}\right] \nonumber \\
&  & + 4P_{4} \left[ k_{2} + k_{2}^{\star} + 3(k_{1}^{\star}+k_{3}) \right]
      + 5 P_{5}\left.-3\eta^{\star}\ell_{4}
        \right\},\\
n_{9} & \equiv & \frac{i}{6\omega}\left\{2P_{2}\left[m_{4}^{\star} +
      m_{8} +  (\ell_{2}^{\star}
      +\ell_{4}) (k_{1}^{\star} + k_{3}\right] + 3P_{3}
      \left[  (k_{1}^{\star} +
       k_{3})^{2} + \ell_{4}^{\star} +
       \ell_{4}\right] \right.\nonumber \\
&    &  + 4P_{4} (k_{1}^{\star}+k_{3}) +  P_{5}\left. \right\}.
\end{eqnarray}

Note that expressions for $P_2$,...,$P_5$ are given in Eq. (A5) in 
Appendix A.
\newpage

\section*{Figure Captions}

\begin{description}
\item[Fig. 1] Plot of $\frac{1}{c_0}$ versus $a$ with $b_0=10^{-4}$. The hatched 
portion shows the instability region which extends between $a=\frac{1}{3}$
and $a=\frac{1}{\sqrt{2}}$ approximately.

\item[Fig. 2] Plots of $\eta_1$  and $\nu_1$ versus $a$ ($b_0=10^{-4}$). 
Note that $\eta_1$ is negative (but finite) in the interval 
$0.4713\le a\le 0.5210$. Similarly, $\nu_1$ is negative (but finite)
in the interval $0.4538\le a\le0.5097$.

\item[Fig. 3] Plots of $|\Psi|^2$ and $P$ versus $a$ in the region of
supercritical bifurcation.

\item[Fig. 4] Plots of $|\Psi|^2$ and $P$ versus $a$ in a subregion of
subcritical bifurcation, i. e. the region where quintic TDGL equation
holds.
 
\item[Fig. 5] Plots of the (numerical) limit cycle solutions obtained from
the secular equations (dotted line), the reduced model (dashed line) and
the $Y^3$-truncated model (crossed line) for $a=0.5$ and $\epsilon=10^{-4}$.

\item[Fig. 6] Plots of the (numerical) limit cycle solutions obtained from
the secular equations (dotted line), the reduced model (dashed line) and
the $Y^5$-truncated model (crossed line) for $a=0.468$ and $\epsilon=10^{-4}$.

\item[Fig. 7] Plots of the (numerical) limit cycle solutions obtained from
the reduced model (dashed line) and the $Y^7$-truncated model (crossed line) 
for $a=0.44$ and $\epsilon=10^{-4}$.

\item[Fig. 8] Plot of the (numerical) limit cycle solution obtained from the
reduced model for $a=0.63$ and $\epsilon=10^{-4}$.

\end{description}


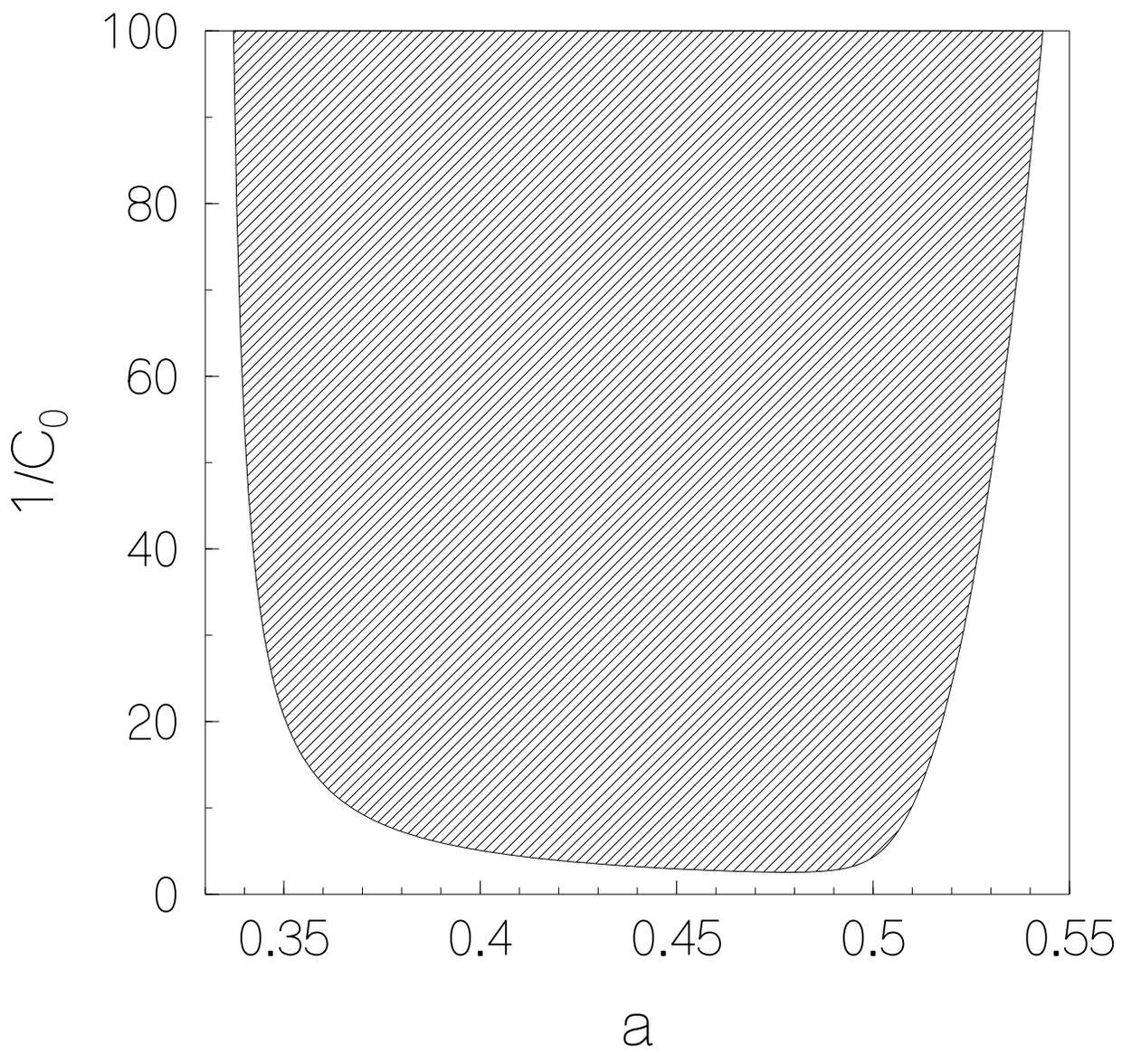
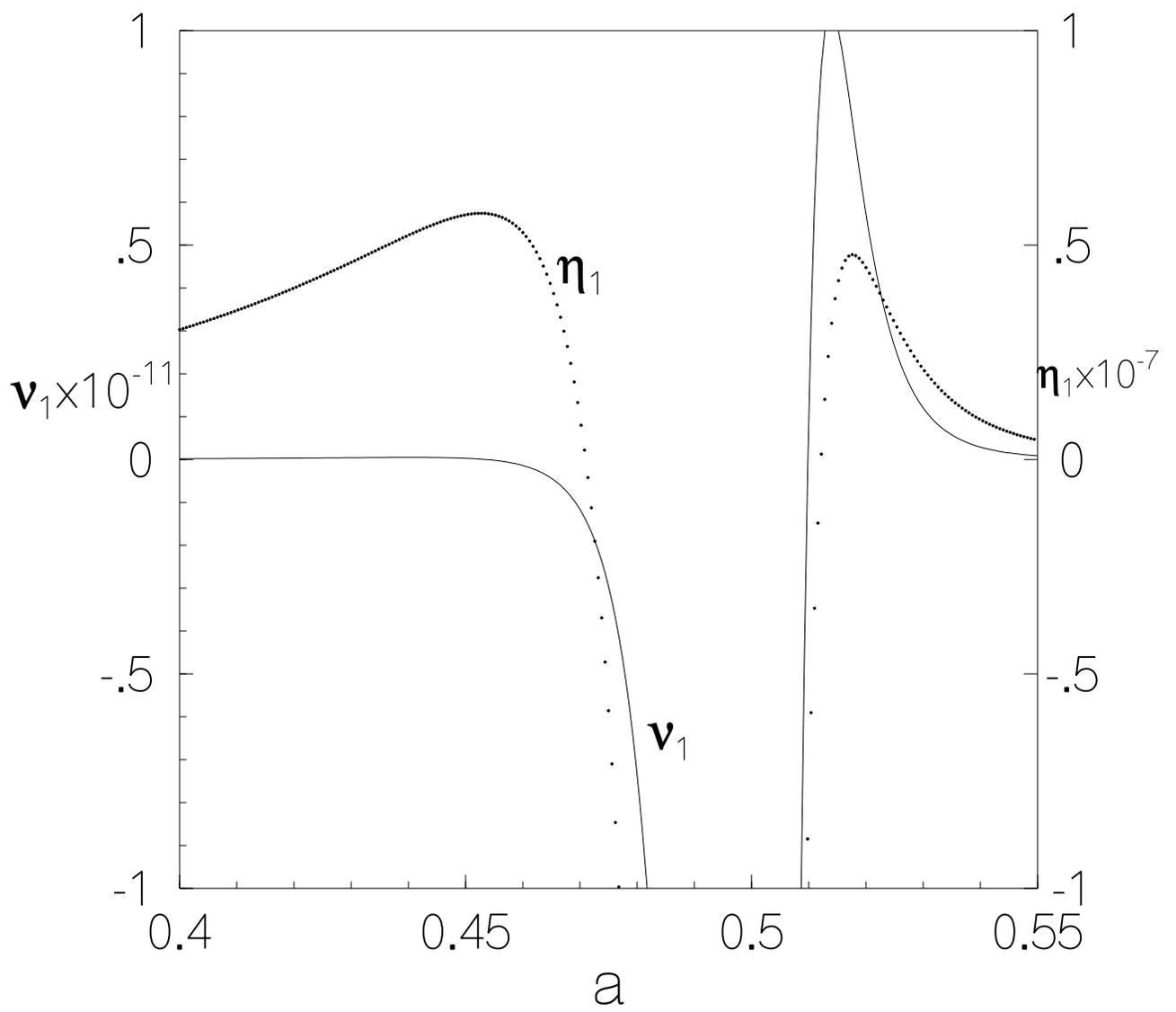
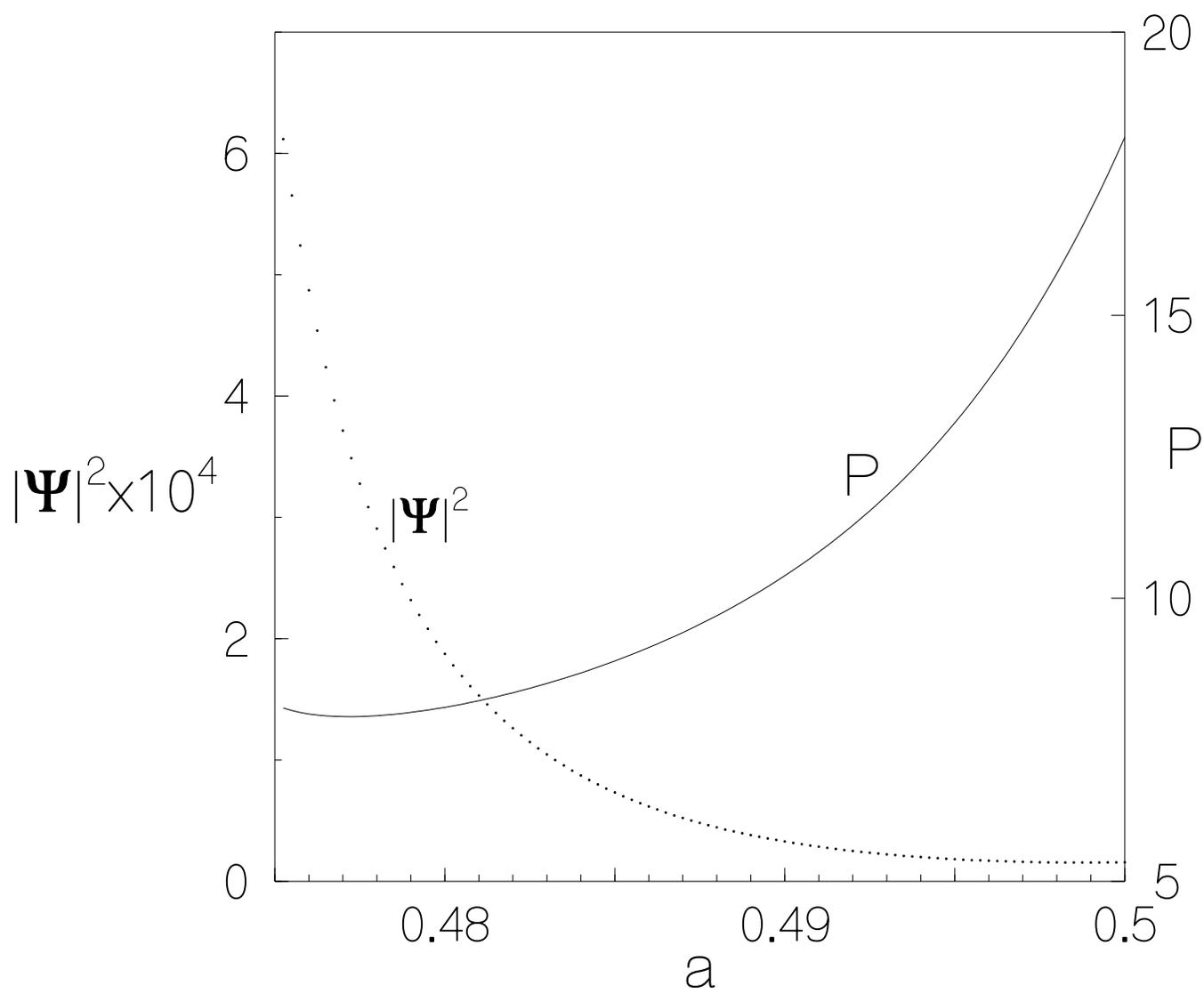
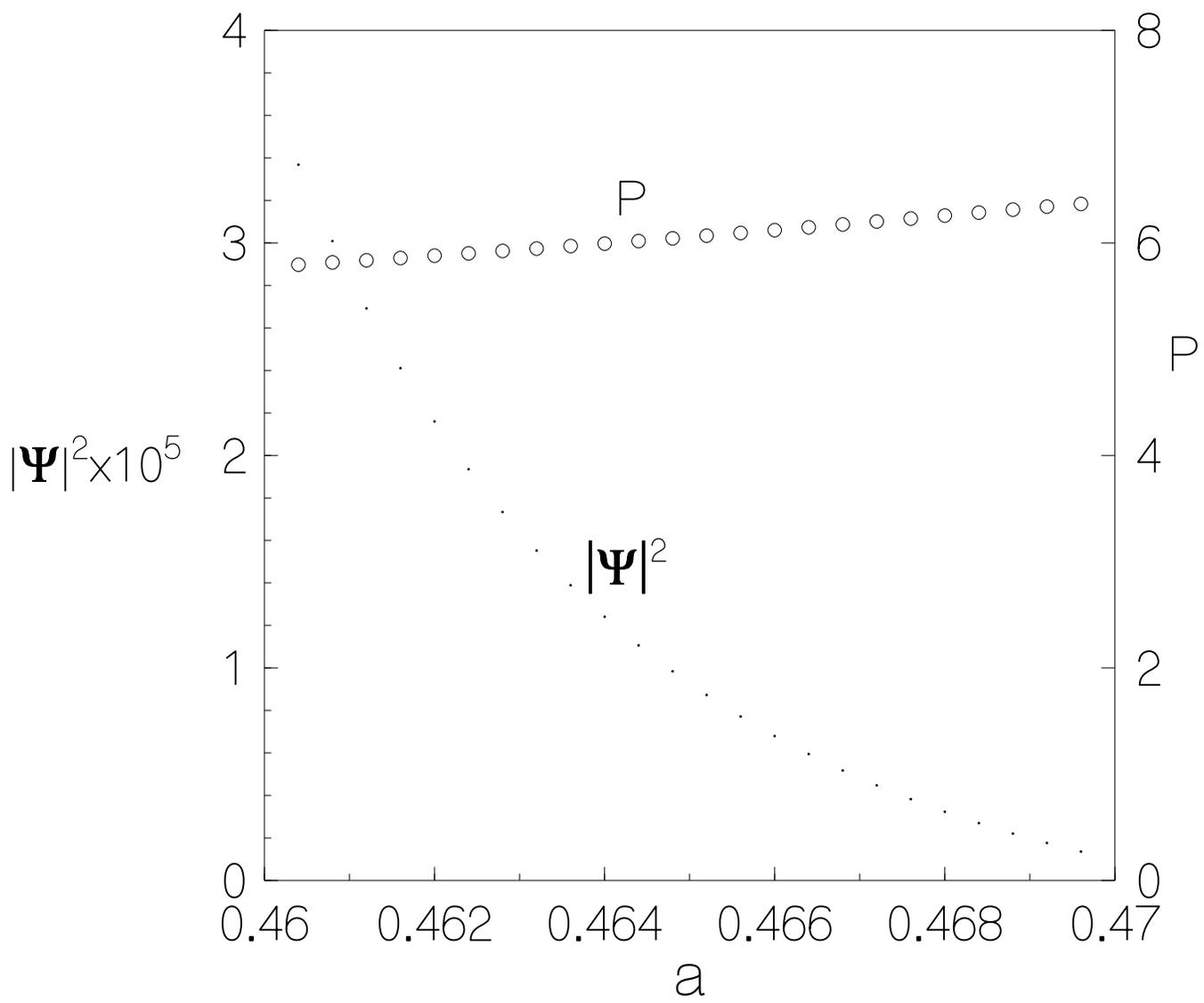
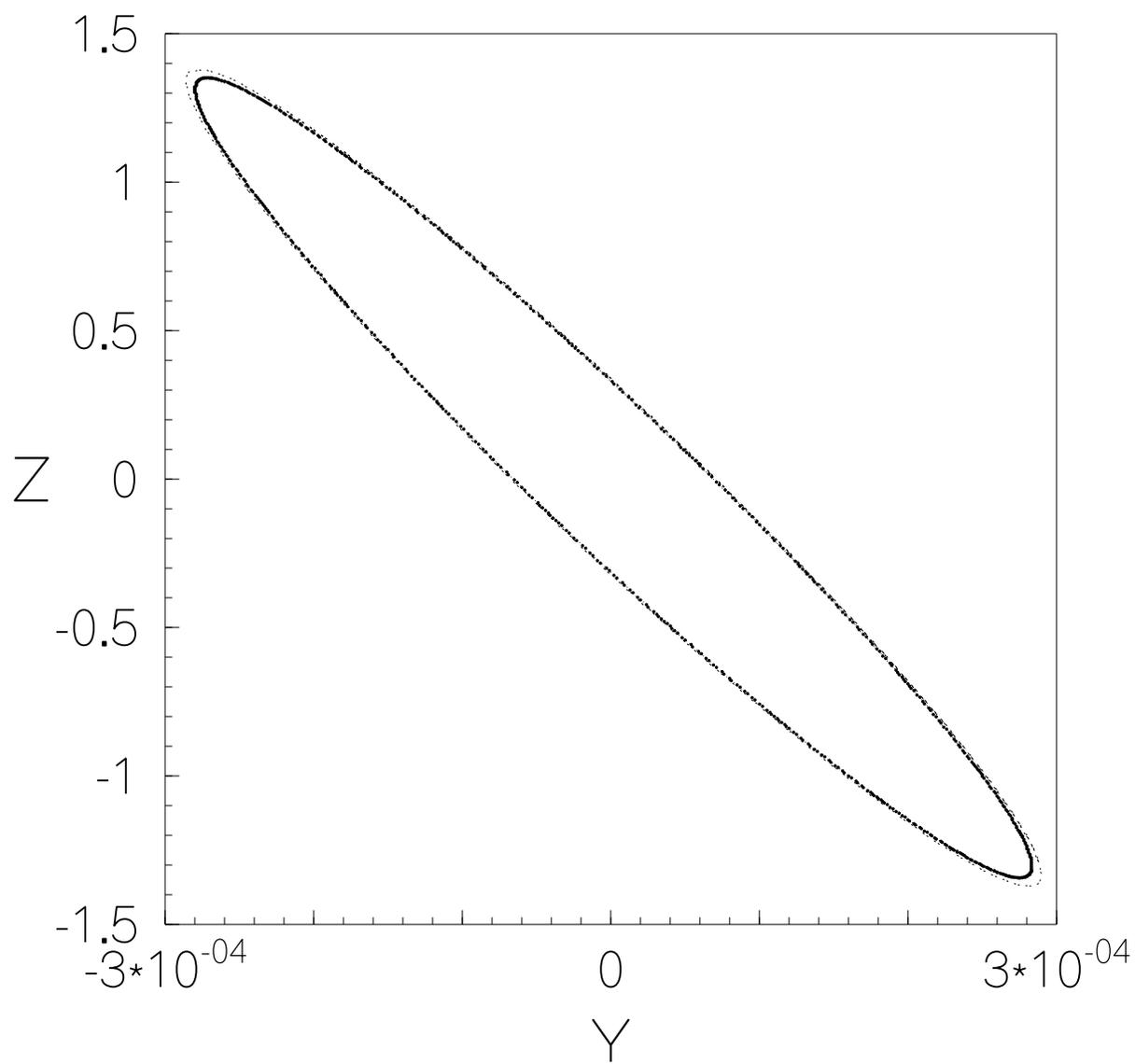
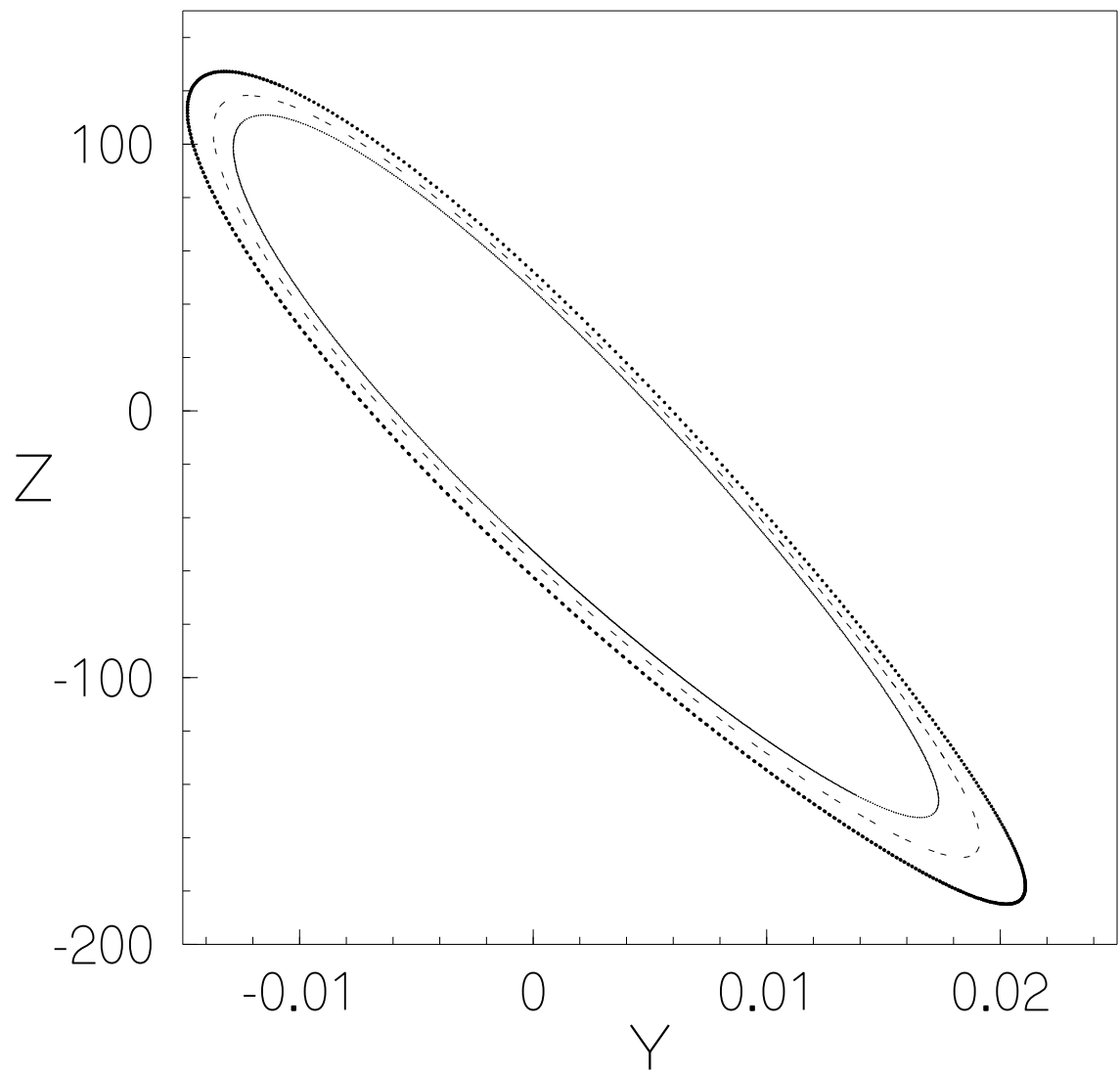
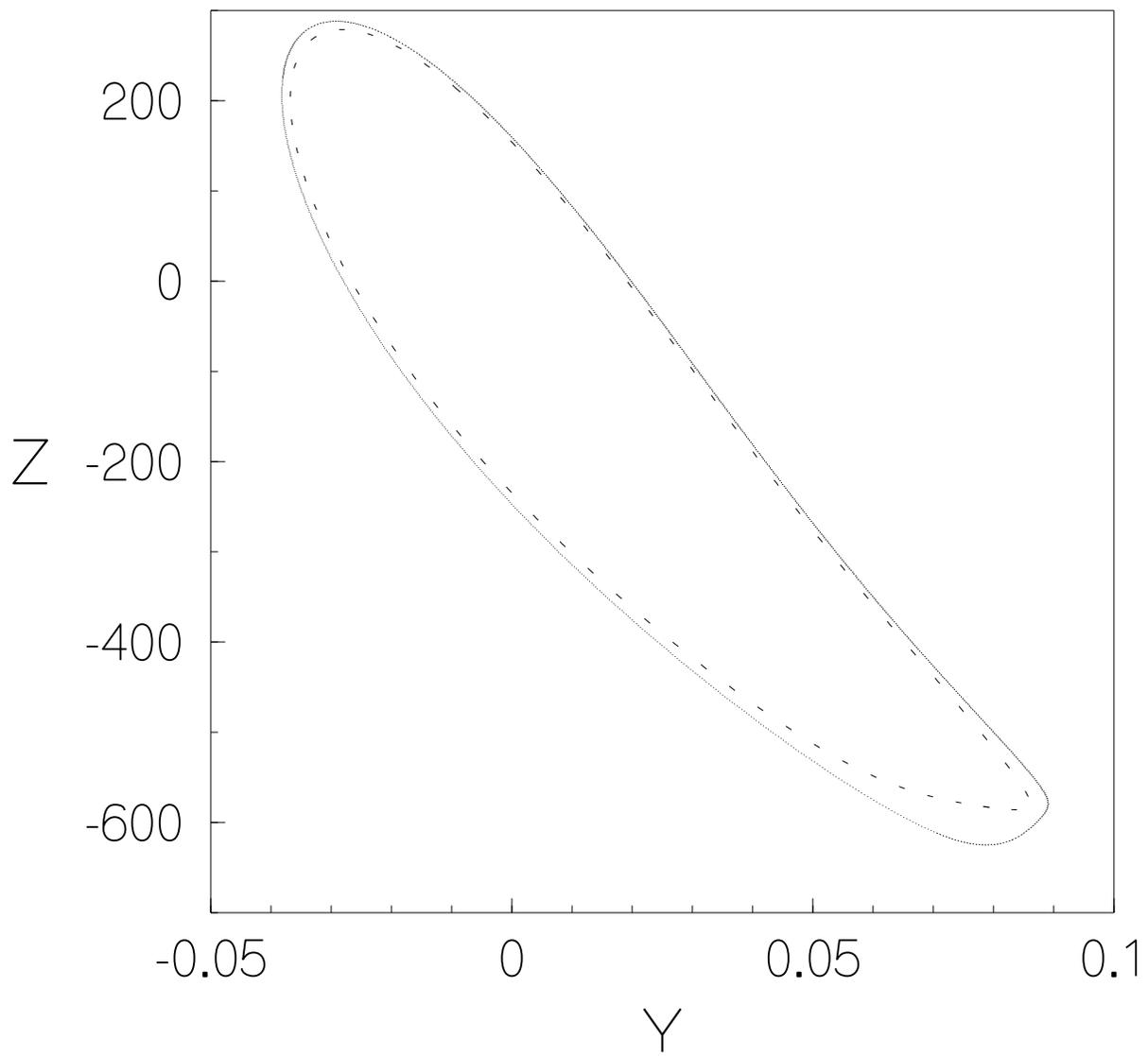
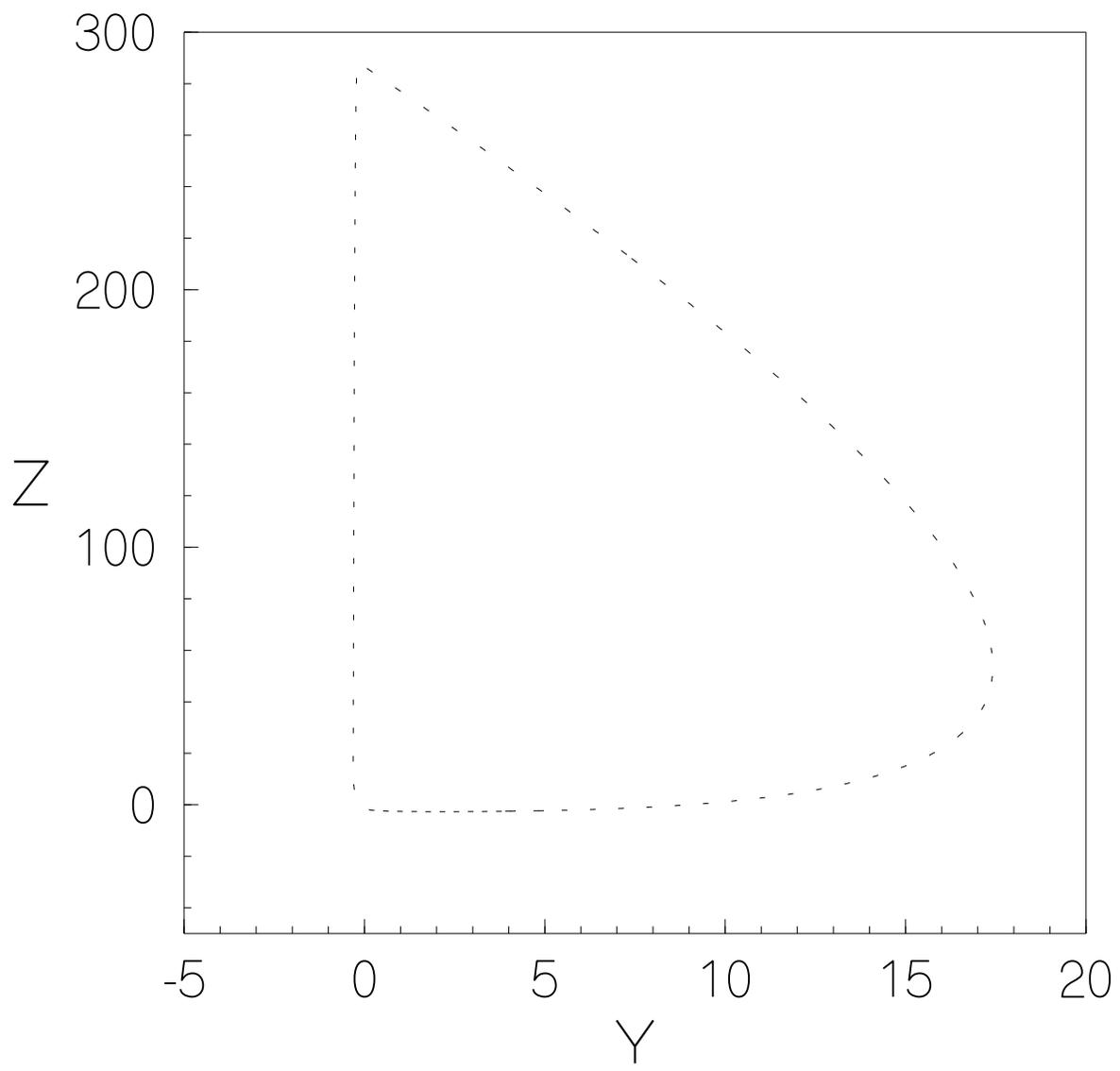
\end{document}